\documentclass[manuscript]{acmart}
\AtBeginDocument{%
  }
\usepackage{longtable,booktabs,array,ragged2e,colortbl,xcolor}
\usepackage{fvextra}
\usepackage{svg}
\setcopyright{acmlicensed}
\copyrightyear{2018}
\acmYear{2018}
\acmDOI{XXXXXXX.XXXXXXX}
\acmConference[Conference acronym 'XX]{Make sure to enter the correct
  conference title from your rights confirmation email}{June 03--05,
  2018}{Woodstock, NY}
  
\acmISBN{978-1-4503-XXXX-X/2018/06}

\begin{document}

\title{A11yLTLNav: Automatic Detection of Accessibility Navigation Failures}

\author{Chenming Ge}
\authornote{Equal contribution.}
\email{gecm@umich.edu}
\orcid{0009-0009-6982-8731}
\correspondingauthor
\authornotemark[1]
\affiliation{%
  \institution{University of Michigan}
  \city{Ann Arbor}
  \state{Michigan}
  \country{USA}
}

\author{Astrid Kewen Peng}
\authornote{Equal contribution.}
\email{talentedastrid@gmail.com}
\orcid{????-????-????-????}
\affiliation{%
  \institution{University of Utah}
  \city{Salt Lake City}
  \state{Utah}
  \country{USA}
}

\author{Chenyang Shi}
\authornote{Equal contribution.}
\email{chengysh@umich.edu}
\orcid{0009-0008-5694-201X}
\affiliation{%
  \institution{University of Michigan}
  \city{Ann Arbor}
  \state{Michigan}
  \country{USA}
}

\author{Ben Greenman}
\email{benjaminlgreenman@gmail.com}
\orcid{0000-0001-7078-9287}
\affiliation{%
  \institution{University of Utah}
  \city{Salt Lake City}
  \state{Utah}
  \country{USA}
}

\author{Yue Jiang}
\email{yue.jiang@utah.edu}
\orcid{0000-0003-0022-6512}
\affiliation{%
  \institution{University of Utah}
  \city{Salt Lake City}
  \state{Utah}
  \country{USA}
}


\begin{abstract} 
  For blind and low-vision (BLV) screen-reader users, a website that appears accessible in a static snapshot can become difficult or impossible to navigate once interaction begins. Yet, most automated accessibility checkers miss failures involving focus, interface state, and accessible feedback across interactions. We present A11yLTLNav, a property-based approach for automatically detecting accessibility navigation failures. 
  Through a structured review of prior research, we organize accessibility navigation failures into a failure taxonomy and formalize a browser-observable subset as executable Linear Temporal Logic properties over action-state traces. A11yLTLNav combines random keyboard exploration with runtime property monitoring to detect these failures during interactions. We evaluate A11yLTLNav on 31 generated websites based on real-world websites and tasks. It reported 309 accessibility failures, of which 274 were confirmed, achieving 88.7\% precision and identifying more confirmed failures than the comparison checkers. Our results show that A11yLTLNav transforms accessibility knowledge into reusable checks of interface behavior over time.
\end{abstract}

\maketitle

\section{Introduction}

The web has become essential to many everyday activities, including accessing information and public services, shopping, travel, education, and employment~\cite{vicente2010multidimensional}. However, accessibility barriers remain common across modern websites~\cite{martins2024large}. These barriers can limit whether people with disabilities can independently access information and complete everyday activities online, especially for blind and low-vision (BLV) users. As essential services increasingly move online, inaccessible navigation can directly restrict BLV users' independence, participation, and access to opportunities.

Prior work has shown that BLV users face difficulties not only in understanding the static content of websites, but also in navigating and interacting with dynamic websites~\cite{borodin2010more,power2012guidelines}. For example, users may lose track of their position after an interface change, have difficulty reaching newly displayed content with the keyboard, or receive insufficient feedback about the result of an action. 
These problems can prevent users from continuing to navigate or completing a task, even when individual elements have appropriate labels or roles. Thus, an interface that appears accessible when inspected in a single state may still become inaccessible as a user navigates through it.

In this work, we define \emph{accessibility navigation failures} as website behaviors that make it difficult for BLV users to reach and operate interactive content, stay oriented as the interface changes, or receive the feedback needed to continue navigating.
Our definition includes both accessibility violations and usability problems when they directly affect BLV users’ experience in navigating and operating a website, following prior work that treats accessibility and usability as distinct but closely related aspects of website use~\cite{petrie2007relationship,aizpurua2016exploring}. 
This definition focuses on the consequences of a failure for navigation rather than only on whether an individual element violates a technical rule.
For example, a dialog may open with appropriate roles and labels while keyboard focus remains on the page behind it. Detecting this failure requires considering the user's activation, the newly opened dialog, and the resulting focus state together.

Prior accessibility research has documented a broad range of accessibility failures encountered by BLV users, including problems with text alternatives, structure and semantics, keyboard and focus, composite widgets, feedback and state, dynamic content, and task navigation~\cite{lazar2007frustrates,borodin2010more,power2012guidelines}. However, knowledge about these failures remains distributed across individual studies and is expressed at different levels of observation, making it difficult to apply systematically in automated testing. We therefore conducted a scoping review of the literature following PRISMA guidelines~\cite{prisma2021statement} to synthesize documented accessibility failures into a unified taxonomy organized by the evidence required to identify them: \textit{element-level} failures can be observed from individual interface elements; \textit{interaction-level} failures emerge across an action and its resulting interface state; and \textit{task-level} failures depend on whether navigation supports a user's broader goal. The taxonomy consolidates prior knowledge into a shared representation and identifies which accessibility navigation failures can be operationalized for automatic detection.

Rule-based accessibility checkers efficiently detect many violations whose evidence is available in a single page state~\cite{vigo2013benchmarking,fischer2025coverage}. Recent checkers extend automated evaluation by driving keyboard interactions~\cite{DBLP:conf/sigsoft/ChiouAH21,chiou2023bagel, chiou2023detecting, jain2025automated}, replaying accessibility tests~\cite{taeb2024axnav}, analyzing screen-reader output~\cite{zhong2025screenaudit}, or reasoning over task execution traces~\cite{zhong2026taskaudit,flow-a11y}. These approaches demonstrate the value of observing interfaces during navigation. However, to our knowledge, no existing accessibility checker translates failure mechanisms documented in prior research into explicit, reusable temporal properties. Consequently, automated checkers either remain limited to predefined static rules or infer accessibility problems from particular execution traces without a reusable specification of the expected behavior. We address this gap by formalizing the browser-observable subset of accessibility navigation failures as Linear Temporal Logic (LTL) properties over action-state traces. An action-state trace records keyboard actions on screen readers and the accessibility-relevant interface states observed before and after them. An LTL property specifies the behavior expected across this sequence, for example, that opening a dialog must be followed by keyboard focus moving inside it. This formalization allows expectations involving focus movement, widget-state changes, and accessible feedback to be evaluated explicitly across interface transitions rather than inferred anew from isolated states or complete task traces.

We propose A11yLTLNav, an LTL-property-based navigation accessibility checker built on Bombadil, a property-based Web testing framework~\cite{bombadil}. 
Property-based testing evaluates interface behavior against explicit properties whenever their triggering conditions occur.
A11yLTLNav performs random keyboard exploration, extracts accessibility-relevant interface states from the websites, and evaluates executable LTL properties over the resulting action-state traces.
Exploration and checking are separated: the explorer determines which keyboard actions on screen readers to execute, while temporal properties determine whether the resulting behavior violates an accessibility expectation. The same property can therefore be checked whenever its trigger arises across different interfaces and exploration paths.
This separation makes each report traceable to a triggering action, an observed outcome, and a violated expectation, while allowing the properties to be reused without defining a complete task for every website.

We evaluate A11yLTLNav on 31 locally hosted website environments generated based on real-world websites and tasks, comparing it with rule-based static accessibility checkers and an agentic accessibility checker adapted from TaskAudit~\cite{zhong2026taskaudit}. A11yLTLNav reported 309 unique accessibility navigation failures, of which 274 were confirmed correct, corresponding to 88.7\% precision. Precision remained between 85.9\% and 88.7\% across three independent exploration runs, while the adapted agentic checker achieved 57.2\% precision. 
A11yLTLNav also identified more confirmed failures than any comparison checker, required 18.52$\times$ less runtime than the agentic checker, and required no language-model inference during testing.

This work makes the following contributions:
\begin{itemize}
\item A unified failure taxonomy that synthesizes accessibility navigation failures documented in prior research and organizes them by the element-, interaction-, or task-level evidence required for detection. Beyond consolidating previously distributed knowledge, the taxonomy identifies the boundary between failures that can be checked from browser-observable behavior and those requiring additional user, task, or assistive-technology context.

\item A11yLTLNav, an automated accessibility checker that operationalizes browser-observable accessibility navigation failures as executable LTL properties and combines state-aware random keyboard exploration with runtime property monitoring over action-state traces.

\item An evaluation on 31 website environments showing that A11yLTLNav detects accessibility navigation failures with high and stable precision, identifies more confirmed failures than the comparison checkers, and operates without language-model inference.

\end{itemize}
\section{Background \& Motivation}

\subsection{Accessibility Barriers in Dynamic Web Navigation}
Blind and low-vision (BLV) users continue to encounter substantial barriers when interacting with modern websites~\cite{WebAIM, vera2025accessible, de2013web, loiacono2009state}. Beyond reducing task efficiency, repeated accessibility failures accumulate into frustration, learned helplessness, and accessibility debt, reflecting a systemic mismatch between mainstream web design and non-visual interaction rather than individual ability~\cite{lazar2007frustrates, khan2026don, harris2011towards, pascual2014impact}. 
These consequences reflect a systemic mismatch between mainstream web design and non-visual navigation rather than a limitation of individual users.
While early text-based websites were comparatively accessible, increasingly dynamic, visually structured interfaces have made accessibility failures more dependent on how users interact with a webpage over time~\cite{lazar2007frustrates, karshmer1995equal, harper2006taming, asakawa2005s}. 

A large body of user studies has documented these failures, including screen-reader navigation problems, focus traps, ARIA misuse, inaccessible widgets, misleading labels, broken jump links, and poorly designed forms~\cite{lazar2007frustrates, power2012guidelines, miao2016contrasting,  cohen2023inaccessible}. Prior work further shows that BLV users may remain unaware of information or functionality that an inaccessible interface never exposes to them~\cite{bigham2017effects}.
Importantly, many of these problems cannot be understood in isolation. The same interface issue may have little impact in one task but completely prevent task completion in another, depending on the interaction history, page state, assistive technology, and user strategy ~\cite{lazar2007frustrates, l2026nonvisual, miao2016contrasting, bigham2017effects}. 
For example, losing focus may be recoverable when the next control is easy to locate but become a blocking failure when an interface update leaves the user without an identifiable navigation position.

Consequently, user studies remain indispensable because they reveal accessibility failures as they unfold during real interaction rather than as isolated interface defects. At the same time, these studies have also revealed rich behavioral phenomena, including navigation strategies, compensatory behaviors, and recurring failure patterns adopted by experienced BLV users ~\cite{borodin2010more}. However, these observations remain fragmented across individual studies and have not been consolidated into a systematic representation that can support automated reasoning.

\subsection{Accessibility Navigation Failures}

Prior work shows that accessibility guideline conformance does not fully reflect the accessibility problems that disabled users experience when interacting with websites. Studies with blind users have found that accessibility and usability problems only partially overlap, and that many problems encountered during real tasks are not covered by accessibility guidelines~\cite{petrie2007relationship,power2012guidelines}. Prior work also found that perceived Web accessibility is closely related to user experience, while its relationship with Web Content Accessibility Guidelines (WCAG) standards is more limited~\cite{aizpurua2016exploring}. Automated evaluation should therefore consider not only whether individual elements satisfy technical requirements, but also whether users can successfully navigate the interface as it changes.

Based on these findings, we define \emph{accessibility navigation failures} as website behaviors that make it difficult for BLV users to reach and operate interactive content, stay oriented as the interface changes, or receive the feedback needed to continue navigating. This definition includes accessibility-conformance violations and usability problems when they directly affect BLV users' ability to navigate and operate a website. For example, a missing label is an accessibility navigation failure because it can prevent a user from identifying a control. A correctly labeled dialog that opens without receiving keyboard focus is also an accessibility navigation failure because the user may be unable to reach its content. Accessible feedback refers to information about an action or state change that is programmatically exposed so that assistive technologies, such as screen readers, can communicate it to the user.
This definition focuses on the consequences of website behavior for navigation rather than treating guideline conformance as the only boundary of accessibility. It also establishes the scope of our failure taxonomy: a problem is included when prior research shows that it can interfere with reaching content, operating controls, maintaining orientation, or receiving the information needed to proceed.

\subsection{Automated Detection and Repair} 
Although user studies are essential for understanding BLV users' experiences, they are resource-intensive and cannot continuously evaluate every page, interaction path, and website update. Automated tools can complement these studies by repeatedly evaluating interfaces at development time and after deployment~\cite{abu2023web}. Static analyzers such as axe~\cite{deque2021axe}, WAVE~\cite{webaim_wave}, and Lighthouse~\cite{googleLighthouse} are highly effective at identifying \textit{element}-level issues, including missing labels, malformed ARIA attributes, and other WCAG violations~\cite{chiou2023bagel, martins2024large}. However, these tools primarily reason over static interface properties, and therefore cannot detect runtime failures that emerge only during interaction ~\cite{borodin2010more, power2012guidelines, miao2016contrasting, WebAIM, loiacono2009state}.

To move beyond static analysis, recent work has explored dynamic accessibility evaluation through scripted interactions, test replay, and interface exploration~\cite{chiou2023bagel,zhong2025screenaudit,fischer2025coverage}. These approaches can expose some interaction- and task-level accessibility navigation failures by observing how an interface responds to user actions. However, scripted approaches generally depend on predefined interaction sequences and checking procedures, limiting their ability to reuse an accessibility expectation whenever the relevant behavior occurs along a different path.

More recently, large language models (LLMs) and Computer Use Agents (CUAs) have been introduced to support accessibility evaluation and repair by executing multistep interface actions and reasoning about task outcomes~\cite{vera2025accessible,a11y-cua,zhong2025screenaudit,lopez2025turning}. A CUA is an agent that observes and operates a graphical interface through actions such as keyboard input, pointer input, or accessibility-service commands. These agents offer greater flexibility than fixed scripts because they can adapt their actions to the current interface and task. However, recent studies show that they remain susceptible to hallucination, execution errors, and inconsistent accessibility judgments~\cite{thisismyfault,khan2026don,perera2026m}. These limitations highlight the need for explicit and inspectable specifications of expected interface behavior rather than reliance on model inference alone~\cite{huq2025automated}.

A11yLTLNav addresses a complementary need: it uses random keyboard exploration to discover interface behaviors and formal temporal properties to determine whether those behaviors satisfy accessibility expectations. This separation preserves the flexibility to encounter failures along different navigation paths while grounding each judgment in an explicit property. Such properties may also provide reliable validation criteria for future repair systems, although the present work focuses on detection.

\subsection{Formal Model for Failure Analysis} 
Formal models provide a principled way to specify how actions, interface states, and expected outcomes relate over time. For example, Stary introduced a user-based modeling method for human-computer interaction~\cite{stary2001user}. By expressing expected sequences of actions, feedback, and state transitions, this decomposition and connection makes it possible to simplify the interaction and scale it into automated analysis~\cite{takagi2007analysis, chiou2023bagel, grillo2014accessible, harper2007web, sarita2025fujma}. It is particularly useful when interactions depends not on a single state, but on how states relate to one another across time ~\cite{o2022quickstrom, chiou2023bagel, flow-a11y, al2016enhancing, watanabe2026aria}. 

Accessibility failures in \textit{interaction}-to-\textit{task} level are inherently temporal. For instance, missing feedback after an action, unexpected focus changes, and unsuccessful recovery from an error are not isolated interface defects. These failures emerge only as interaction sequences unfold. Although prior BLV accessibility research has documented these failures through user studies, barrier-based evaluations, and qualitative reports~\cite{chiou2023bagel,zhong2025screenaudit,power2012guidelines}, the resulting accessibility knowledge has not been systematically represented as reusable properties for runtime checking.

We address this gap by synthesizing documented accessibility navigation failures into a unified failure taxonomy and translating the browser-observable subset into Linear Temporal Logic (LTL) properties. LTL is a formal language for specifying how conditions should hold across an ordered sequence of states. It can express, for example, that a condition must always hold, must eventually become true, or must become true after a particular event~\cite{porfirio2018authoring,lee2025veriplan}. In A11yLTLNav, these properties are evaluated over \emph{action-state traces}: ordered records of keyboard actions and the accessibility-relevant interface states observed before and after them.

Each property defines a triggering condition and an expected outcome. For example, when a keyboard action opens a modal dialog, the resulting state should place keyboard focus inside that dialog. A violation of this expectation provides concrete evidence of an accessibility navigation failure. This representation makes expected and prohibited interface behaviors explicit, executable, and reusable across websites. Temporal logic therefore bridges accessibility knowledge documented in prior research and scalable automated evaluation, enabling A11yLTLNav to check not only what an interface contains, but also how it behaves as users navigate.

\section{Failure Taxonomy}
To support systematic analysis and formal reasoning, we synthesize accessibility navigation failures documented in prior research into a unified failure taxonomy. Following the user-centered definition of accessibility~\cite{petrie2007relationship}, we include both accessibility violations and usability problems when they make it difficult for BLV users to reach or operate interactive content, stay oriented as an interface changes, or receive the feedback needed to continue navigating.

\begin{table}[t]
\centering
\caption{Comparison of three levels of BLV web accessibility evaluation. Our work derives task-level ecological validity from empirical evidence while preserving the scalability of formal checking.}
\label{tab:eval-levels}
\small
\setlength{\tabcolsep}{5pt}
\renewcommand{\arraystretch}{1.3}
\begin{tabular}{@{}p{2.5cm}p{2.0cm}p{3.5cm}p{2.7cm}p{2.7cm}@{}}
\toprule
\textbf{Taxonomy Level} & \textbf{Failure Target} & \textbf{Detection Methods} & \textbf{Scalability} & \textbf{Ecological Validity} \\
\midrule
\textbf{\textit{L1: Element-level}}
& Elements, Labels, ARIA
& \textbf{Rule-based Detection} e.g., axe~\cite{deque2021axe}, WAVE~\cite{webaim_wave}, Lighthouse~\cite{googleLighthouse}
& Automatic \newline \textit{(no human required)} & Low \newline \textit{(static properties only)} \\
\addlinespace[0.2em]
\textbf{\textit{L2: Interaction-level}}
& Interaction Steps
& \textbf{Simulated interaction:} barrier walkthrough ~\cite{brajnik2008comparative}, expert usability evaluations~\cite{miao2016contrasting}
& Semi-automatic \newline \textit{(requires expert setup)} & Partial \newline \textit{(expert-simulated, not naturalistic)} \\
\addlinespace[0.2em]
\textbf{\textit{L3: Task-level}}
& Real-world tasks
& \textbf{Real-user observation:} user studies~\cite{bigham2007webinsitu,huq2026bridging, pereira2015preliminary, leuthold2008beyond}, think-aloud~\cite{borodin2010more}
& Manual \newline \textit{(requires real users)} & High \newline  \textit{(naturalistic tasks \& context)} \\
\midrule
\textbf{Ours} \newline \textit{L2 + L3}
& Interactive failure patterns from real-world tasks
& \textbf{Formalized Observations Detection}
& \textbf{Automatic} \newline \textit{(temporal property checking)}
& \textbf{High} \newline \textit{(derived from real-world observations)}\\
\bottomrule
\end{tabular}
\end{table}

\subsection{Scoping Review Method}

Our work focuses on bridging the gap between \textit{scalability} and \textit{ecological validity} for accessibility navigation failures that extend beyond isolated interface elements. Instead of relying only on isolated conformance violations~\cite{chiou2023bagel}, we synthesize interaction- and task-level accessibility navigation failures into a unified taxonomy that serves as the foundation for formal, automated reasoning.

We constructed the taxonomy through a structured multi-source literature review following PRISMA ~\cite{prisma2021statement, prisma2021explanation}. We search four databases (ACM Digital Library, IEEE, Springer and Science Direct) for research articles using complementary queries over abstract fields: population (e.g., "blind" "low vision") × context (e.g., "web") × outcome (e.g., "failure""usability") query in July, 2026. Specifically, ACM Digital Library yields $196$ papers (190 via abstract searching+ 6 via title); IEEE Xplore yields $138$ papers; Springer yields $383$ papers(originally 33,967 papers, narrowing to Article, and HCI subjects  "Interaction design", "Web accessibility evaluation and user experience", "User interfaces and human computer interaction", Subdisciplines of "User interfaces and human computer interaction" and Language of "English"); ScienceDirect $238$ papers. To assess the recall of the primary query,  $6$ additional papers were retained from two supplementary searches in August, 2026: Q1 applied broader term combinations (n = 115) and Q2 targeted keyboard- and interaction-level failure terminology (n = 16), deduplication excluded. Detailed queries attached in the Appendix. $145$ papers were removed due to deplication and wrong content type (e.g., magazines, proposals,etc.). Of the $936$ candidate papers, $460$ were excluded in the screening stage for "Irrelevant Topics" (SC1), "Not Full Paper"(SC2), and "Non-English"(SC3). SC1 involves mobile apps and softwares, independent games and blind studies. $476$ met our inclusion criteria of reporting empirically observed interaction failures encountered by BLV or keyboard-only users on web platforms. Failures extracted from the final $56$ studies were iteratively coded, consolidated, and refined into $13$ taxonomy categories organized by the stage of interaction in which they occur. In addition to database searches, we conducted one round of backward snowballing on all papers retained after full-text screening, examining their reference lists for relevant works not captured by keyword search. This yielded $4$ additional candidates.

\subsection{Levels of Accessibility Failure}

Based on the review, we found that existing accessibility research has progressively expanded from evaluating isolated interface elements to understanding user interaction and real-world task performance. We organize prior work into three complementary levels of accessibility evaluation: conformance-level (L1), interaction-level (L2), and task-level (L3). \autoref{tab:eval-levels} summarizes the targets, common detection methods, scalability, and ecological validity associated with these three levels and situates our approach in relation to them.

\textit{\textbf{L1: Element-level}} evaluation asks whether individual interface elements satisfy accessibility rules and can be flagged by rule-based tools and checklists~\cite{brajnik2008comparative, thomsen2006building}. This includes WCAG success criteria~\cite{prakash2026micro, brajnik2008comparative, costa2015differences}, whose A/AA/AAA levels indicate the classification of a given criterion, national or regional guidelines, and other specific rules derived from experts~\cite{zhong2025screenaudit, fukuda2005proposing}. Such approaches are automatic and effective at identifying node-level issues, including missing labels, malformed ARIA attributes, and insufficient contrast. However, they primarily reason over static interface properties and cannot determine whether these issues ultimately prevent successful navigation~\cite{asakawa2005s}.

\textit{\textbf{L2: Interaction-level}} evaluation shifts the focus from interface elements to interaction steps, asking whether users can successfully navigate and operate an interface. Representative approaches include barrier walkthroughs, expert usability evaluations, and interaction-based accessibility testing~\cite{brajnik2008comparative, miao2016contrasting, asakawa2005s}. Rather than checking isolated violations, these simulation methods reveal failures such as navigation confusion, missing feedback, focus loss, and interaction dead ends that emerge during interaction. Recent systems such as BAGEL further extend automated evaluation from isolated interface elements to interaction flow by assessing keyboard navigability and interaction sequences~\cite{chiou2023bagel}. Nevertheless, these approaches still evaluate whether individual interaction steps remain executable under predefined evaluation procedures rather than whether users ultimately accomplish real-world tasks in their natural contexts~\cite{brajnik2009guideline, asakawa2005s}.

\textit{\textbf{L3: Task-level}} evaluation asks whether a sequence of interactions ultimately supports successful completion of users' real-world goals. Real-user observation studies demonstrate that the same interaction issue may be recoverable in one task but prevent task completion in another, depending on the user's strategy, interaction history, assistive technology, and task context~\cite{borodin2010more,takagi2007analysis,huq2026bridging, saulynas2022putting}. Such studies offer high ecological validity, revealing recurring navigation strategies, compensatory behaviors, and context-dependent failure patterns that cannot be explained by isolated interaction steps alone ~\cite{bigham2007webinsitu, pereira2015preliminary, leuthold2008beyond, asakawa2005s}. However, these findings remain largely qualitative and fragmented across individual studies, making them difficult to scale for automated reasoning.

\autoref{tab:taxonomy} and \autoref{tab:taxonomy3} present the resulting failure taxonomy for L2 and L3 (The full failure taxonomy for L1, L2, and L3 is in the Appendix), including each accessibility navigation failure, an illustrative example, and the prior studies documenting it. By organizing accessibility navigation failures according to their required evidence, the taxonomy identifies which failures can be evaluated from browser-observable interface behavior and therefore provides the basis for their subsequent formalization as temporal properties. 

\begin{table}[p]
\caption{Taxonomy of Level 2 BLV web accessibility failures, organized by the level of
evidence required to observe them. Citations after each failure give the
reporting studies.}\label{tab:taxonomy}
\small
\renewcommand{\arraystretch}{1.2}
\begin{tabular}{@{}p{1.5cm} p{6cm} p{7.2cm}@{}}
\toprule
\textbf{Category} & \textbf{Failure} & \textbf{Example}\\
\midrule

\multicolumn{3}{@{}l}{\textbf{L2: Interaction-level}}\\
\midrule
Keyboard and focus & Loss of focus after activation \cite{a11y-cua,harper2003middleware,fakrudeen2025evaluation,pakdeechote2012new,harper2000pilot,huq2026bridging} & after confirming a departure airport, focus returns to the top of the page and the date fields must be found again\\
 & No exit from a composite widget \cite{a11y-cua} & Tab is intercepted inside a video description panel and never advances to the next control\\
 & Improperly hidden elements remain focusable \cite{a11y-cua,wentz2011usability,whitelaw2003make} & Tab stops on items in a collapsed carousel and the screen reader announces an empty string\\
Composite widgets & No valid active option in expanded combobox \cite{a11y-cua,wentz2011usability} & the suggestion list is open but no option is marked as current, so nothing is announced while arrowing\\
 & No active-option movement on arrow keys \cite{a11y-cua,whitelaw2003make} & pressing Down inside an open suggestion list does not advance to the next entry\\
 & No commit state after option selection \cite{a11y-cua,wentz2011usability} & Enter on a highlighted airport leaves the field showing a partial string, and the user believes the choice was taken\\
 & No valid active cell in a managed grid \cite{a11y-cua} & a date picker opens with no cell marked as current, so arrow keys have no starting point\\
 & No active-cell movement on arrow keys \cite{a11y-cua} & Tab passes over the whole calendar, and the arrow keys that would move within it are never announced\\
 & No effect from a role-specific activation key \cite{a11y-cua,pascual2014impact} & a calendar cell is selected only by Space; pressing Enter repeatedly produces nothing\\
Feedback and state & No feedback confirming an activation \cite{buzzi2010facebook,leporini2011google,huq2026bridging,power2012guidelines,ryskeldiev2022investigating,wentz2011usability,borodin2008s} & after submitting a form nothing is announced, and the user re-reads the page to find out whether it was accepted\\
 & Misplaced or disconnected response to an activation \cite{sandnes2024consent,harper2000pilot,uckun2020breaking,borodin2010more,a11y-cua,wentz2011usability,borodin2008s,whitelaw2003make} & a filter updates a results panel further down the page with nothing tying the two together; correcting a typo re-opens the suggestion list on every keystroke\\
 & ARIA state contradicts the native state \cite{a11y-cua} & a checkbox reported as \texttt{aria-checked="false"} while the native control is checked\\
 & No exposed expanded state after a reveal \cite{a11y-cua,asakawa2005s} & an accordion panel opens visually but the trigger continues to report itself as collapsed\\
 & No trace of operation progress or mode \cite{huq2026bridging,li2026content,wentz2011usability} & an upload gives no sign of whether it is running or has stalled; the same key both inspects and edits a block, with nothing saying which just happened\\
Dynamic content & No accessible name or named dismissal control on a dialog \cite{gupta2005extracting,wentz2011separate,leporini2011google,csontos2021accessibility,a11y-cua,borodin2010more} & a consent banner is announced only as \emph{dialog}, with no way to tell what it is or how to close it\\
 & No focus move into a newly opened dialog \cite{a11y-cua,borodin2010more,wentz2011usability} & a cookie notice appears over the page while focus stays behind it, so the user keeps reading content that is now blocked\\
 & No dialog dismissal or focus return on Escape \cite{a11y-cua} & Escape does not close the overlay; when it finally closes, focus is left at the top rather than on the control that opened it\\
 & No reachable control to start playing or stop playing media \cite{tomlinson2016perceptions,miyashita2007making,pascual2014impact} & a video begins on load and its pause control is neither reachable by Tab nor announced\\
Structure \& semantics & Inadequate labels to reveal nearby context \cite{buzzi2010facebook,wentz2011usability,borodin2008s,asakawa2005s,whitelaw2003make,leuthold2008beyond} & a price is read out with no indication of which product row or which column it belongs to\\
AT interoperability & Inconsistent name exposure across reading modes \cite{buzzi2010facebook,potluri2021examining} & a password field is announced by name while browsing, but only as \emph{edit box} once the user starts typing\\
 & Screen-reader shortcut captured outside page \cite{a11y-cua,makati2024promise} & the context-menu command opens the browser's own menu instead of the page's; an injected accessibility widget claims keys the user's screen reader needs\\
\end{tabular}
\end{table}

\begin{table}[t]
\caption{Taxonomy of Level 3 BLV web accessibility failures.}\label{tab:taxonomy3}
\small
\renewcommand{\arraystretch}{1.05}
\begin{tabular}{@{}p{1.5cm} p{6cm} p{7.2cm}@{}}
\multicolumn{3}{@{}l}{\textbf{L3: Task-level}}\\
\midrule
Feedback and state & No guidance for recovering from a rejected action \cite{huq2026bridging,ryskeldiev2022investigating,borodin2008s} & a submission is refused with no indication of which field was wrong or what format it expected; a category must exist before an expense can be added, and nothing says so\\
Text alternatives & Inadequate description of goal-relevant information \cite{ferreira2012aligning,power2012guidelines,yu2025cluttered,fukuda2005proposing,harper2003middleware,alam2023seechart,loiacono2009state,l2026nonvisual,yesilada2003rendering,krejtz2025higher,uckun2020breaking,ryskeldiev2022investigating,islam2010mixture,borodin2010more,leporini2011google,kodandaram2026finding,gleason2020twitter,jeong2023wataa,bigham2006webinsight,whitelaw2003make,pascual2014impact,leuthold2008beyond} & a binary tree described as running prose, leaving parent and child relations unrecoverable; a chart summarised in one sentence when the task requires comparing series; a link named \emph{read more} among twenty others\\
Structure \& semantics & Reading order and grouping inconsistent with task's logic \cite{yu2025cluttered,gadde2014screen,potluri2021examining,leporini2011google,ryskeldiev2022investigating,krejtz2025higher,pakdeechote2012new,power2012guidelines,ferreira2012aligning,aizpurua2015prejudices,wentz2011separate,yang2013bypassing} & a product's price is read far from its name; the store-locator sits under \emph{Account} rather than under \emph{Stores}\\
 & Excessive heading marks relative to page content \cite{yu2025cluttered,gadde2014screen} & every list item carries a heading, so the heading list is as long as the page\\
 & Tab order inconsistent with task's order \cite{fakrudeen2025evaluation,asakawa2005s} & the submit button is reached before the fields it depends on\\
 & Position learned from prior visits no longer predicts where the control is \cite{huq2026bridging,wentz2011usability} & the submit button sits in the top bar on one page and after the last field on the next\\
Navigation efficiency & No efficient route to the control the task requires \cite{fukuda2005proposing,ferdous2023enabling,power2012guidelines,da2018content,potluri2021examining,a11y-cua,ryskeldiev2022investigating,borodin2010more,wentz2011usability} & reaching a store-selection setting takes several hundred Tab presses through the footer\\
 & No confirmation of completeness, driving re-verification \cite{gadde2014screen,kodandaram2026finding,harper2003middleware,borodin2010more,bigham2017effects,borodin2008s} & a forum thread is re-read from the top because nothing confirms that no later reply revised the answer\\
 & No way to resume from a prior position \cite{rohani2012back,prakash2023autodesc,borodin2008s} & returning from a product page drops the user at the top of the results list rather than at the item they left\\
\end{tabular}
\end{table}

\section{A11yLTLNav: A Failure-driven Checker for Navigation Accessibility Failures}

\begin{figure*}[t]
    \centering
    \includegraphics[width=\textwidth]{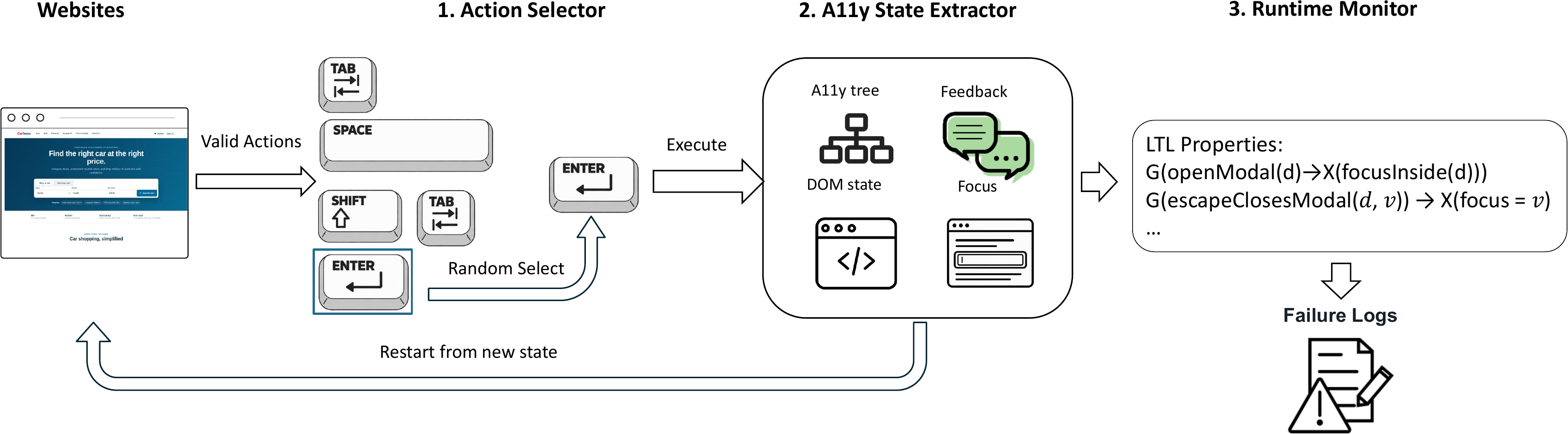}
    \caption{
    Overview of our temporal property testing workflow.
    From each webpage state, the action selector chooses a valid keyboard
    action. The accessibility state extractor captures the resulting interface
    state including a11y tree, DOM and ARIA state, focus and accessible feedback, which is checked against executable temporal properties by the
    runtime monitor. Violations are recorded as failure logs, while exploration
    continues from the new state.
    }
    \Description{A left-to-right diagram shows a keyboard exploration loop with three numbered components. A webpage supplies valid actions to the action selector, which executes a keyboard action. The accessibility state extractor records the accessibility tree, document object model state, focus, and feedback. These observations feed a runtime monitor, whose example rule requires focus to enter a newly opened modal in the next relevant state. Detected violations produce failure logs. A return arrow from the state extractor to the webpage indicates that exploration restarts from the newly reached state.}
    \label{fig:system-overview}
\end{figure*}

To automatically detect accessibility navigation failures in websites, we built A11yLTLNav, an accessibility checker that automatically detects these failures. Figure \ref{fig:system-overview} shows the overall pipeline of our checker. 

\subsection{Temporal Property-Based Testing for Accessibility Navigation Failures}

We model an interaction as a trace of user actions and the interface states that change following these actions. Many accessibility navigation failures are not identifiable from a single interface state. For example, a failure occurs when activating a button opens a modal but keyboard focus remains in the background page, which cannot be detected through static analysis of the elements. Therefore, detecting such failures requires modeling the relationship between an interaction event and the interface states that follow it.

We use Linear Temporal Logic (LTL) to represent these interaction expectations over execution traces. LTL provides temporal operators for conditions that should always hold ($\mathbf{G}$), hold after an interaction ($\mathbf{X}$), or eventually become true ($\mathbf{F}$).
Appendix~\ref{appendix::ltl} presents a brief formal semantics of LTL.
These operators allow us to express an accessibility navigation failure as an interaction trigger followed by an expected state transition. We use LTL to formalize generic, interaction level properties that determine whether an observed interface transition is accessible for BLV users. None of our LTL formulas require knowledge of the user's website-specific task goal.

\subsection{Property-Based Checker for Accessibility Navigation Failures}

Existing accessibility checkers typically evaluate a page state at the moment the checker is invoked~\cite{deque2021axe, webaim_wave, zhong2025screenaudit}. In contrast, navigation failures may occur only in the process of state transactions.  Traditional browser automation frameworks such as Playwright~\cite{playwright} provide the frameworks needed to perform keyboard actions and inspect browser state, but are primarily based on example-based tests where the test author specifies the assertions and elements that should be evaluated. Such tests work well for known workflows, but do not easily generalize the same interaction check to controls and state changes that have not been identified in advance, since would require specifying the interaction paths where each accessibility failure may occur.

We therefore build A11yLTLNav on Bombadil~\cite{bombadil}, a property-based testing framework that provides state-aware random exploration and temporal property checking for building our checker. For exploration, Bombadil repeatedly captures the current browser state, constructs the actions that are available in that state, and randomly selects an action to execute. Because the available actions are generated from the current interface state, this exploration can check interaction sequences that are not specified in advance. The same accessibility property can therefore be checked across different interface states whenever its triggering condition arises. For checking, Bombadil allows developers to define correctness properties over the sequence of browser states produced during exploration. Its property-based language directly supports temporal operators, allowing the LTL properties derived from our failure taxonomy to be implemented as executable scripts. Bombadil thus provides both the random exploration mechanism used to expose candidate interaction failures and the temporal checking infrastructure used to detect violations.

However, Bombadil is still a general-purpose UI testing framework rather than an accessibility checker. Its default interaction space and browser observations are not designed around keyboard navigation, focus management, or accessibility-relevant interface state. We therefore extend Bombadil with three components: an \emph{action selector}, an \emph{accessibility state extractor}, and a \emph{property monitor}.

\subsubsection{Action Selector} Bombadil randomly select actions from sighted users, like click and scroll, to explore the websites. BLV users instead use keyboard-only and screen reader actions. Therefore, we replace Bombadil's default action space with a state-aware keyboard action generator tailored to BLV users' navigation methods. At each browser state, the generator determines the valid action set based on the currently focused element and widget context: 
\begin{itemize}
    \item Sequential navigation (\texttt{Tab}, \texttt{Shift+Tab}\footnote{Since Bombadil do not support \texttt{Shift+Tab}, we modified the action generator to support this action.}) is universally enabled. 
    \item Activation actions (\texttt{Enter}, \texttt{Space}) are exposed for interactive controls.
    \item Directional navigation (arrow keys) is enabled in composite widgets like comboboxes, grids, tabs, and menus.
    \item Espace actions (\texttt{Escape}) are enabled in dialogs or layered overlays.
\end{itemize}

During exploration, Bombadil selects actions from this action space.  Exploration and property checking are decoupled: the explorer decides where and how to navigate, while runtime monitors independently verify whether the observed trace violates an accessibility property. The exploration continues even after detecting a defect, allowing a single session to discover multiple distinct violation witnesses within its time budget.

\subsubsection{Accessibility State Extractor}

Bombadil provides the browser exploration and temporal-property execution infrastructure, but its browser state does not directly expose the accessibility-relevant information required by our properties. We therefore extend Bombadil with an accessibility state layer that derives four types of observations from each browser state: \emph{accessibility semantics}, \emph{DOM and ARIA state}, \emph{focus}, and \emph{interaction feedback}. Accessibility semantics capture information relevant to the accessibility tree, such as element roles and accessible names derived from native marks and ARIA attributes. DOM and ARIA state captures interaction-relevant properties such as whether dialogs or controlled regions are open or visible, and values including \texttt{aria-expanded}, \texttt{aria-checked}, and related state attributes. Focus observations record the currently focused element and changes in focus across interactions. Finally, feedback observations capture changes that can communicate the result of an action, including updates to live regions, status messages, and alerts. 

\subsubsection{Runtime Property Monitor.}

During exploration, A11yLTLNav continuously evaluates the written LTL properties against the browser states produced by the action selector. The monitor creates a property-specific transaction when a trusted keyboard action satisfies the property's triggering condition. The transaction binds the triggering action to the relevant interface object and records the pre-action observations required by that property. At the subsequent settled browser state, the monitor collects the corresponding post-action evidence and determines whether the property is applicable, whether sufficient evidence is available to evaluate it, and whether the observed transition violates the expected behavior. An observation is reported as a violation only when the property is applicable.

\subsection{From Empirical Failures to Executable Temporal Properties}

\subsubsection{From Failure Taxonomy to Temporal Properties}

Our failure taxonomy captures accessibility navigation failures observed across prior BLV studies. We turn these empirical observations into temporal properties by identifying three components for each failure taxonomy: the interaction that triggers the relevant behavior, the interface outcome expected after that interaction, and the observations needed to determine whether the expectation holds. This process converts qualitative descriptions of accessibility failures into reusable temporal properties that can be checked automatically across different websites and interaction paths. For failure taxonomy items that span multiple cases, we formulate several temporal properties.

\begin{figure*}[t]
    \centering
    \includegraphics[width=1\textwidth]
        {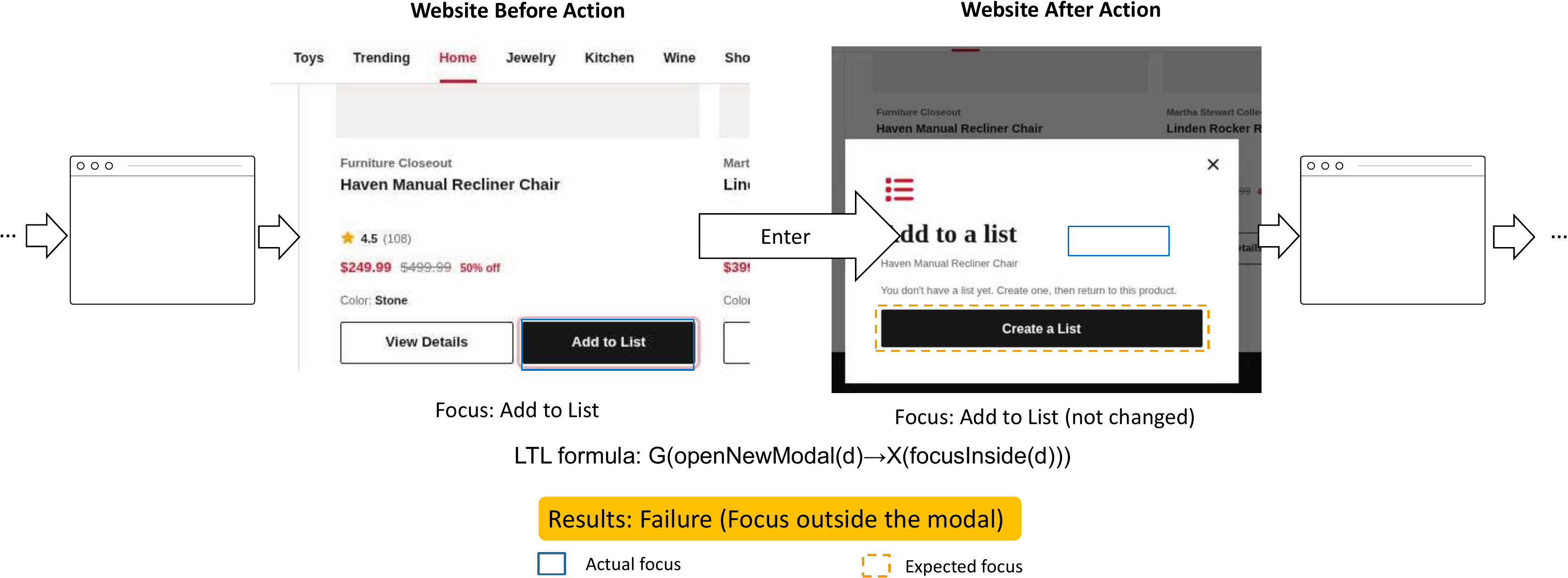}
    \caption{
        Example of a focus-management failure.
        Activating ``Add to List'' opens a dialog, but focus remains on the trigger.
    }
    \Description{Two screenshots show keyboard focus before and after opening an Add to List dialog. Initially, the Add to List button on a product listing has focus. Pressing Enter opens an overlay containing an Add to a list heading and a Create a List button, while focus remains on the original button behind the overlay. The temporal rule requires focus to be inside any newly opened modal in the next relevant state. The result is labeled Failure because the observed focus remains outside the modal.}
    \label{fig:property-example}
\end{figure*}

For example, as shown in Figure~\ref{fig:property-example}, our taxonomy includes cases in which opening a modal does not move keyboard focus into the new interaction context. This failure corresponds to a violation of the following LTL property, where $d$ denotes the newly opened modal dialog:
$
\mathbf{G}\bigl(
    \mathrm{openModal}(d)
    \rightarrow
    \mathbf{X}\,\mathrm{focusInside}(d)
\bigr)
$
The property states that whenever an interaction opens a new modal, keyboard focus should enter that same modal in the next relevant post-action state. This formulation captures the temporal relationship between the triggering interaction and its expected accessibility outcome, rather than checking either interface state in isolation.

\begin{figure*}[t]
    \centering
    \includegraphics[width=\linewidth]{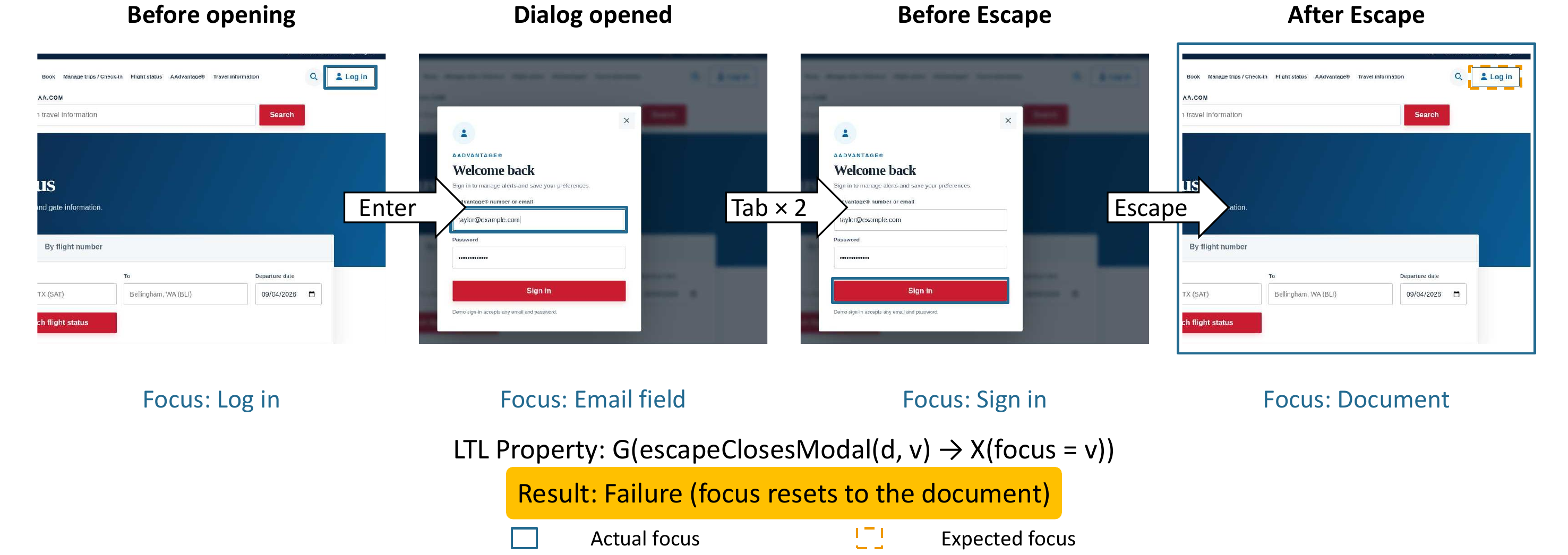}
    \caption{
Example of a focus-return
failure after dialog closes. Activating \emph{Log in} opens the
dialog and establishes the button as its invoker. Escape closes the dialog, but focus resets to the
document instead of returning to the invoker.
}
\Description{Four screenshots follow focus through opening and dismissing a login dialog. From left to right, focus starts on Log in; Enter opens the dialog and moves focus to the email field; two Tab presses move focus to Sign in; and Escape closes the dialog but leaves focus on the document. Solid outlines mark actual focus. A dashed outline marks the expected final focus on the original Log in button. The temporal rule requires an Escape-dismissed modal to return focus to its recorded invoker in the next relevant state. The illustrated sequence violates this rule.}
    \label{fig:invoker-focus-return}
\end{figure*}

Figure~\ref{fig:invoker-focus-return} illustrates another property when a modal closes. Pressing Enter on Log in opens a dialog and moves focus into the email field. Two Tab actions then move focus to Sign in. Pressing Escape closes the dialog, but focus resets to the document instead of returning to Log in. We express the expected focus return as follows:
\(
\mathbf{G}\bigl(
    \mathrm{escapeClosesModal}(d,v)
    \rightarrow
    \mathbf{X}(\mathrm{focus}=v)
\bigr)\).
Here, \(d\) denotes the modal and \(v\) denotes its invoker, the button used to open it. The triggering condition captures an Escape action that closes the modal while its invoker remains eligible to receive focus. The property requires focus to return to that same button in the next relevant post-action state.

Checking this expectation requires retaining information across the interaction sequence: the checker records the invoker when the modal opens and compares it with the observed focus after the modal closes. In this example, Log in remains available, but the resulting focus is on the document, violating the property. The expected destination therefore depends on how the user entered the dialog, not just on the page state after dismissal.

The same failure category may produce multiple temporal properties when different interaction mechanisms expose distinct observable behaviors. In this way, the taxonomy provides the empirical foundation, while the temporal properties provide precise and reusable guidelines for automated checking. The complete failure-to-property mapping is provided in Appendix~\ref{appendix:failure-property-mapping}.

\subsubsection{From Temporal Properties to Executable Scripts}

Afterwards, we convert each temporal property into a runtime script to monitor over the interaction trace. Each script identifies the triggering interaction, binds it to the relevant state extracted by previous steps, records the required observations before the action, and evaluates the expected behavior after the interface settles with the action.
This formulation allows the property to capture a common accessibility expectation without prescribing a single implementation mechanism. Different interfaces may communicate the same successful action through different accessible signals, while the monitor checks whether the resulting transition remains perceivable through accessibility-relevant browser state.

%

For example, in Figure~\ref{fig:invoker-focus-return}, the script records Log in as the invoker when Enter opens the dialog. Moving focus to the email field and then Sign in does not change this association. After Escape, the script observes that the dialog has closed and Log in remains eligible, but focus is on the document. These observations establish the violation of the property.

\section{Evaluation}

We evaluate A11yLTLNav's performance by comparing to several state-of-the-art accessibility checkers: axe, WAVE and taskAudit~\cite{deque2021axe, webaim_wave, zhong2026taskaudit}.
We study the following research questions:

\textbf{RQ1}: How does A11yLTLNav compare with existing accessibility checkers in precision, and how stable is its precision across exploration runs?

\textbf{RQ2}: How quickly does A11yLTLNav check the accessibility failures comparaed to other accessibility checkers?

\textbf{RQ3}: How do A11yLTLNav and other accessibility checkers differ in the failures they detect and diagnose?

We evaluate the tools on 31 websites. For static checkers, we ran them on the first page of each website and excluded failures that are not targeted for BLV users. For the agentic checker, we ran the pipeline using DeepSeek V4 Flash 0731~\cite{xu2026deepseek}.  For A11yLTLNav, we run three trials of 30-minutes each, because the exploration process is random.
We ran all experiments on Rocky Linux 8.10, allocating one Intel Xeon Gold 6530 CPU core and 4GB of memory to each run.

\subsection{Baselines}

\emph{Rule-based static accessibility scanner}. Axe \cite{deque2021axe} and WAVE \cite{webaim_wave} are representative rule-based accessibility checkers commonly used for automated web accessibility evaluation. These tools inspect the webpage and identify violations of static accessibility rules.

\emph{Task-driven agentic checker}. We implement a task-driven agentic accessibility checker following TaskAudit~\cite{zhong2026taskaudit}, the closest prior approach to our setting since it targets accessibility failures through the process of task execution. For each generated website, a web agent powered by browser-use \cite{browser_use2024} attempts each task using screen reader transcript as observations and keyboard and screen reader actions for navigation. The interaction traces are then analyzed for accessibility failures using an LLM. Our prompts to the agentic checker are shown in Appendix~\ref{appendix::prompttaskAudit}.

\subsection{Website Generation}

It is challenging to evaluate live websites using automatic tools because the websites are designed for humans, and often block tools~\cite{salehnamadi2023assistive, taeb2024axnav, huq2026bridging, anupam2025browserarena}.
Also, since website behavior changes over time, live websites are a poor setting for reproducible experiments.
Following prior work that reconstructs real websites as executable synthetic environments \cite{gao2026training, chae2026safe, zhou2026webarena, zhang2026infiniteweb}, we evaluate our checker on locally hosted generated websites. We generated a subset of cloned websites from Mind2Web~\cite{deng2023mind2web}. Following CowCorpus's method for website selection \cite{huq2026modeling}, for each of Mind2Web's 31 subdomains, we selected the website with the highest U.S. traffic according to Similarweb, resulting in 31 websites. We synthesized these environments using Claude Code \cite{claudecode} with GPT-5.6 Sol \cite{openai2026gpt56sol}, adapting the website generation pipeline from uxCUA \cite{gao2026training}. The prompt for website generation is attached in Appendix~\ref{appendix::promptuxCUA}.

\subsection{Metrics}

We evaluate each checker by the precision of its reported accessibility failures and its efficiency in terms of execution time and cost. Firstly, we deduplicated the detected failures. Two reported accessibility failures were grouped when they belonged to the same component of the same website and were reported by the same checker. Then, we followed the definition of the accessibility failures discussed before when annotating these reported failures. Two of the authors manually reviewed all reported failures across checkers against the corresponding interaction traces and the error definitions in our failure taxonomy. Each report of accessibility errors by these checkers was annotated for whether it represented a correct detection or a false positive, and afterwards we resolved different opinions through discussion.

\section{Results}

Table~\ref{tab:correctness} summarizes the precision of reported accessibility failures from all four checkers. Because axe and WAVE primarily evaluate page-level failures and find fewer failures then others, our analysis for the result focus on the task-driven agentic checker and A11yLTLNav, both of which observe accessibility failures during runtime interactions.

\subsection{RQ1: Precision and Stability}

\subsubsection{Agentic Checker}

The task-driven agentic checker reported 276 failures across 31 websites, of which 158 were confirmed correct and 118 were false positives, with a precision of 57.2\% (95\% CI: 52.0--62.5\%). Its main advantage is that it has access to both task context and screen-reader output, which allows it to reason about failures whose significance depends on what the user is trying to accomplish.

However, its reports of accessibility failures also depend on correctly executing and interpreting the task trace. Among 516 task runs by the web agent executor, 306 (59.3\%) reached the maximum step limit and 367 (71.1\%) contained at least one recorded action error. Our review further found false positives where the analyzer relied on an assumed website state from the screen reader output rather than the actual focus and action outcome from the websites, or inferred missing information from a single screen-reader output even though later output provided it. Moreover, the agents' thoughts and traces are different from that of BLV users in the process of task execution, resulting in different subgoals and trajectories in the process of task execution~\cite{a11y-cua}. Thus, task context provides useful semantic evidence, but also introduces uncertainty from both agent execution and LLM-based trace interpretation.

\subsubsection{A11yLTLNav}

A11yLTLNav reported 309 accessibility failures, of which 274 were confirmed correct and 35 were false positives, corresponding to a precision of 88.7\% (95\% CI: 82.2--93.8\%). This was 31.4\% higher than the agentic checker (95\% CI: 23.2--39.3\%).
%
%
The false positives were primarily associated with the boundary of particular properties, such as imperfect conditions or accessibility failure properties not represented by the current state extractor. We examine representative correct and incorrect detections in the case analysis below.


\begin{table}[t]
  \centering
  \caption{Precision of static checkers, the agentic checker, and A11yLTLNav
  on deduplicated reports. The shaded row denotes our method; boldface
  indicates the best result for each evaluation metric.}
  \label{tab:correctness}
  \small
  \setlength{\tabcolsep}{4pt}
  \begin{tabular}{@{}lrrrr@{}}
  \toprule
  \textbf{Checker} &
  \textbf{Reports} &
  \textbf{Correct} &
  \textbf{False Positives} &
  \textbf{Precision} \\
  \midrule
  Axe             & 18  & 17  & \textbf{1}   & \textbf{94.4\%} \\
  WAVE            & 68  & 34  & 34            & 50.0\% \\
  Agentic checker & 276 & 158 & 118           & 57.2\% \\
  \midrule
  \rowcolor{black!7}
  \textbf{A11yLTLNav (ours)}
                  & 309 & \textbf{274} & 35 & 88.7\% \\
  \bottomrule
  \end{tabular}
\end{table}

\emph{Stability}.
Because A11yLTLNav's exploration is random, we additionally check precision across the three independent runs. Precision remained between 85.9\% and 88.7\% across runs, shown in Table~\ref{tab:replay-correctness}, suggesting that although different exploration runs expose different interactions, the reported violations remain consistently precise. The overall row represents reports consolidated accessibility errors across runs.

 \begin{table}[t]
  \centering
  \caption{Review of A11yLTLNav reports for accessibility errors across three exploration runs.}
  \label{tab:replay-correctness}
  \small
  \begin{tabular}{@{}lrrrr@{}}
  \toprule
  \textbf{Run} &
  \textbf{Total} &
  \textbf{Correct} &
  \textbf{False Positive} &
  \textbf{Precision} \\
  \midrule
  Run 1 & 309 & \textbf{274} & 35 &\textbf{ 88.7\%} \\
  Run 2 & 313 & 269 & 44 & 85.9\% \\
  Run 3 & 299 & 257 & 42 & 86.0\% \\
  \midrule
  Pass@3   & 493 & \textbf{424} & 69 & 86.0\% \\
  \bottomrule
  \end{tabular}
  \end{table}

\subsection{RQ2: Runtime and Computational Cost}

The task-driven agentic checker required substantially more computation than A11yLTLNav. Across the 31 websites, TaskAudit required 288.20 hours in total. This corresponds to an average runtime of 557.81 minutes per website (SD = 247.86 minutes). In comparison, A11yLTLNav completed exploration and property checking for Run~1 in 15.56 hours, averaging 30.12 minutes per website (SD = 0.01 minutes). Overall, the agentic checker required approximately $18.52\times$ as much runtime as
A11yLTLNav.

\begin{table}[t]
  \centering
  \caption{Runtime comparison across 31 websites. The shaded row denotes our method; boldface indicates the lower runtime.}
  \label{tab:runtime}
  \small
  \setlength{\tabcolsep}{6pt}
  \begin{tabular}{@{}lrr@{}}
    \toprule
    \textbf{Checker} &
    \textbf{Total Runtime (h)} &
    \textbf{Mean / Website (min)} \\
    \midrule
    Agentic checker & 288.20 & 557.81 \\
    \midrule
    \rowcolor{black!7}
    \textbf{A11yLTLNav (ours)} & \textbf{15.56} & \textbf{30.12} \\
    \bottomrule
  \end{tabular}
\end{table}

The agentic checker also incurred substantial language-model usage. Across task execution and accessibility analysis, Agentic checker consumed approximately 1.01 billion tokens, including 945.38 million prompt tokens and 60.30 million output tokens, corresponding to \$63.11 in API cost. For each website, Agentic checker consumed an average of 30.50 M input tokens (SD = 12.03 M) and 1.95 M output tokens (SD = 0.81 M), for an average of 32.44 million tokens in total (SD = 12.74 M). At the August 2026 DeepSeek V4 Flash pricing, the resulting average API cost was \$2.04 per website (SD = \$0.80). In contrast, A11yLTLNav does not invoke an LLM during exploration or property evaluation and therefore does not have any token cost.


\begin{table*}[t]
  \centering
  \caption{Token usage and estimated cost across 31 websites. A11yLTLNav uses zero tokens.}
  \label{tab:agent-token-cost}
  \small
  \setlength{\tabcolsep}{7pt}
  \begin{tabular}{@{}llrrrr@{}}
    \toprule
    \textbf{Checker} &
    \textbf{Statistic} &
    \textbf{Prompt Input (M tokens)} &
    \textbf{Output (M tokens)} &
    \textbf{Total (M tokens)} &
    \textbf{Cost (USD)} \\
    \midrule
    {Agentic checker}
      & Total
      & 945.38 & 60.30 & 1005.69 & 63.11 \\
      & Per-site mean
      & 30.50 & 1.95 & 32.44 & 2.04 \\
    \bottomrule
  \end{tabular}
\end{table*}

\subsection{RQ3: Comparative Analysis of Detected Failures}

\begin{figure*}[t]
\centering
\includegraphics[width=\textwidth]{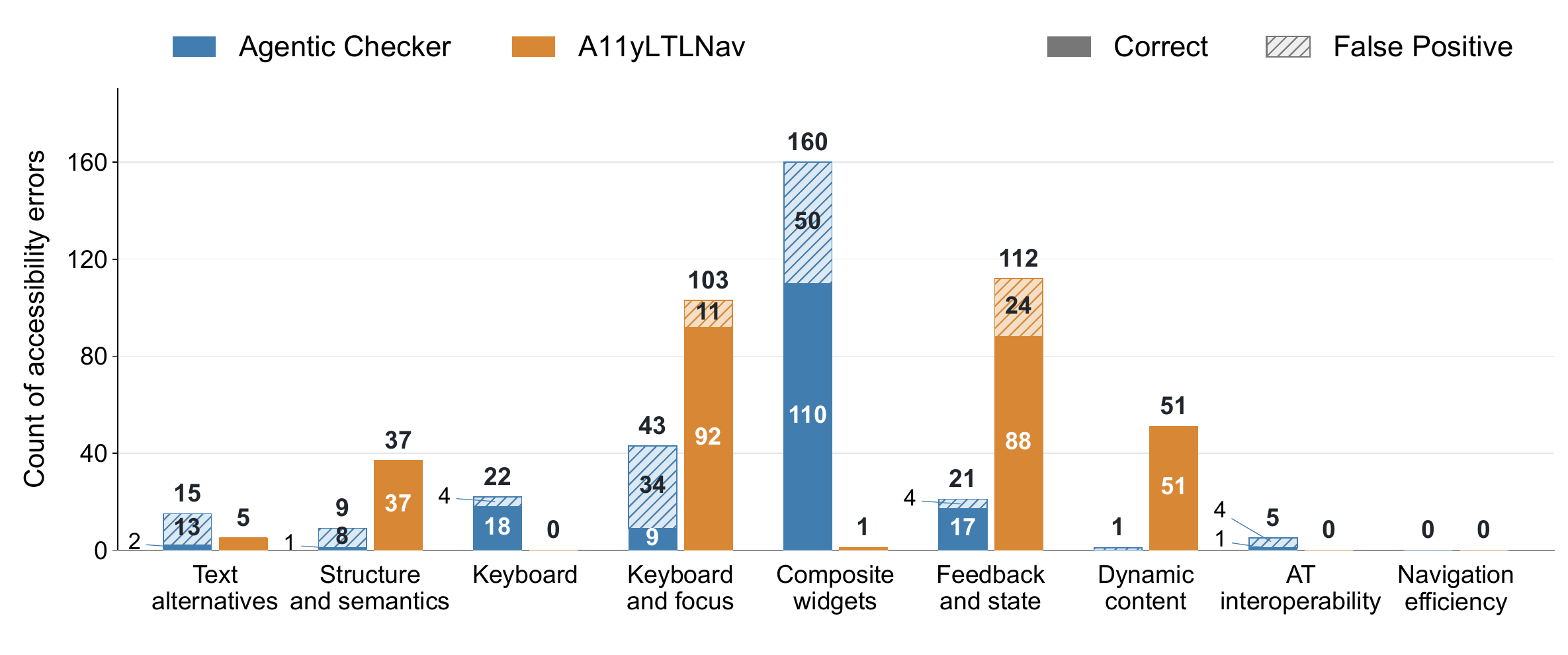}
\caption{Accessibility error reports grouped by failure domain from the failure taxonomy. }
\Description{A grouped stacked bar chart compares reported findings across nine failure domains. The horizontal axis lists domains and the vertical axis gives report counts, with ticks from zero to 160. Within each pair, the left bar is the agentic checker and the right bar is A11yLTLNav. Solid segments represent correct reports and hatched segments represent false positives. Total reports for the agentic checker and A11yLTLNav, respectively, are: text alternatives, 15 and 5; structure and semantics, 9 and 37; keyboard, 22 and 0; keyboard and focus, 43 and 103; composite widgets, 160 and 1; feedback and state, 21 and 112; dynamic content, 1 and 51; assistive-technology interoperability, 5 and 0; and navigation efficiency, 0 and 0. Of the agentic checker's 160 composite-widget reports, 110 are correct and 50 are false positives. A11yLTLNav has 92 correct and 11 false-positive keyboard-and-focus reports, and 88 correct and 24 false-positive feedback-and-state reports. These are counts of reported findings, not recall measurements.}
\label{fig:taxonomy-domain-reports}
\end{figure*}

To further examine RQ3, Figure~\ref{fig:taxonomy-domain-reports} groups the reports from the two runtime checkers by failure domain in our taxonomy. We use the evidence provided in the reports of accessibility errors to aggregate domains across taxonomy levels because the same interaction concern can arise with different levels of evidence. 

The two approaches produced substantially different failure reports in the figure. The agentic checker was dominated by reports involving \emph{composite widgets}, which accounted for 160 of its 276 reports, followed by \emph{keyboard and focus} (43), \emph{keyboard} (22), and \emph{feedback and state} (21). In contrast, A11yLTLNav's reports were concentrated in \emph{feedback and state} (112), \emph{keyboard and focus} (103), and \emph{dynamic content} (51). It additionally reported 37 \emph{structure and semantics} failures.

These distributions reflect the different evidence and checking boundaries of the two approaches. A11yLTLNav's largest domains correspond to generalizable failures that can often be expressed as consistent relationships between a keyboard action and a browser-observable outcome, such as whether focus remains valid, an interface state is exposed consistently, or accessible feedback follows an activation. Composite-widget behavior is more different across widgets, websites and tasks: different widgets support different interaction models, states, and keyboard conventions, making them less easy to be captured by a small reusable property set. The agentic checker can instead interpret these interactions from task and screen-reader context without requiring a predefined property for each observable mechanism. These results therefore suggest different detection profiles by either approach; the following cases illustrate the mechanisms behind these differences.

\subsubsection{A11yLTLNav}

\begin{figure}[t]
    \centering
    \includegraphics[width=\linewidth]{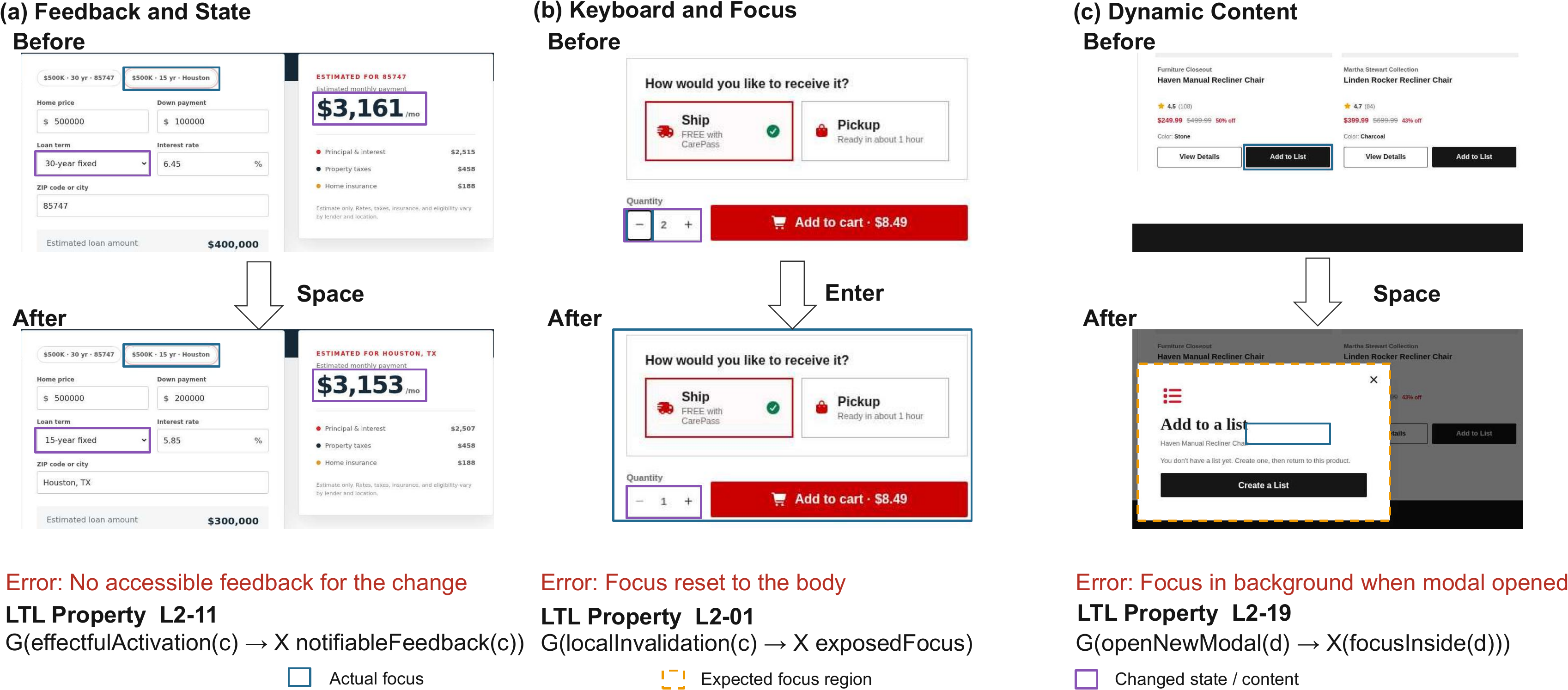}
    \caption{Examples of accessibility navigation errors captured by A11yLTLNav and their corresponding temporal properties. Each example shows the interface state before and after a keyboard action, together with the property that specifies the expected accessible state transition: (a) missing accessible feedback after an effectful activation, (b) focus loss after a local interface update, and (c) focus remaining outside a newly opened modal. Colored outlines indicate observed focus, expected focus regions, and changed interface state or content.}
    \Description{Three panels pair before-and-after screenshots with keyboard actions and temporal rules. In panel a, pressing Space on a mortgage preset changes form values and the monthly estimate from 3,161 to 3,153 dollars, but provides none of the accessible feedback signals monitored by property L2-11. In panel b, pressing Enter on the quantity-decrease control changes the quantity from two to one but resets focus to the document; property L2-01 requires exposed focus after the local update. In panel c, pressing Space on Add to List opens a modal while focus stays on the background button; property L2-19 requires focus inside the new modal. Outlines identify actual focus, the expected focus region, and changed values or content.}
    \label{fig:name-temporal-examples}
\end{figure}

The three A11yLTLNav examples in Figure \ref{fig:name-temporal-examples} illustrate how temporal properties connect a keyboard action to an expected interface state. In each case, the interface visibly responds, but the resulting state does not satisfy the accessibility expectation.

For \emph{feedback and state}, pressing Space on a setting of the loan updates the form values and monthly estimate. However, the update produces none of the accessible confirmation signals monitored by A11yLTLNav. Script of the LTL property L2-11 checks that an accessible activation is followed by notifiable feedback. The report therefore distinguishes a successful update from the missing confirmation accompanying it.

For \emph{keyboard and focus}, pressing Enter on a button to decrease quantity changes the quantity from two to one. However, keyboard focus then resets to the body of the page, leaving no control focused. Script of the LTL property L2-01 checks that focus remains exposed after a local change invalidates the focused control. The failure concerns losing focus during a small update within the same view.

For \emph{dynamic content}, pressing Space on Add to List opens a dialog, but keyboard focus remains on the background button behind it. Property L2-19 checks that opening a new modal is followed by focus inside it. The dialog's appearance alone does not establish that keyboard users can interact with the newly displayed content.  

Together, these examples illustrate why A11yLTLNav can detect a broad set of interaction-level accessibility failures while maintaining high precision. By summarizing each judgment in an temporal property between a triggering action and its expected accessibility outcome, A11yLTLNav can systematically detect failures across many interactions while reducing errors from ambiguous or incomplete interpretation of execution traces.

\subsubsection{Agentic Checker}

\begin{figure}[t]
    \centering
    \includegraphics[width=\linewidth]{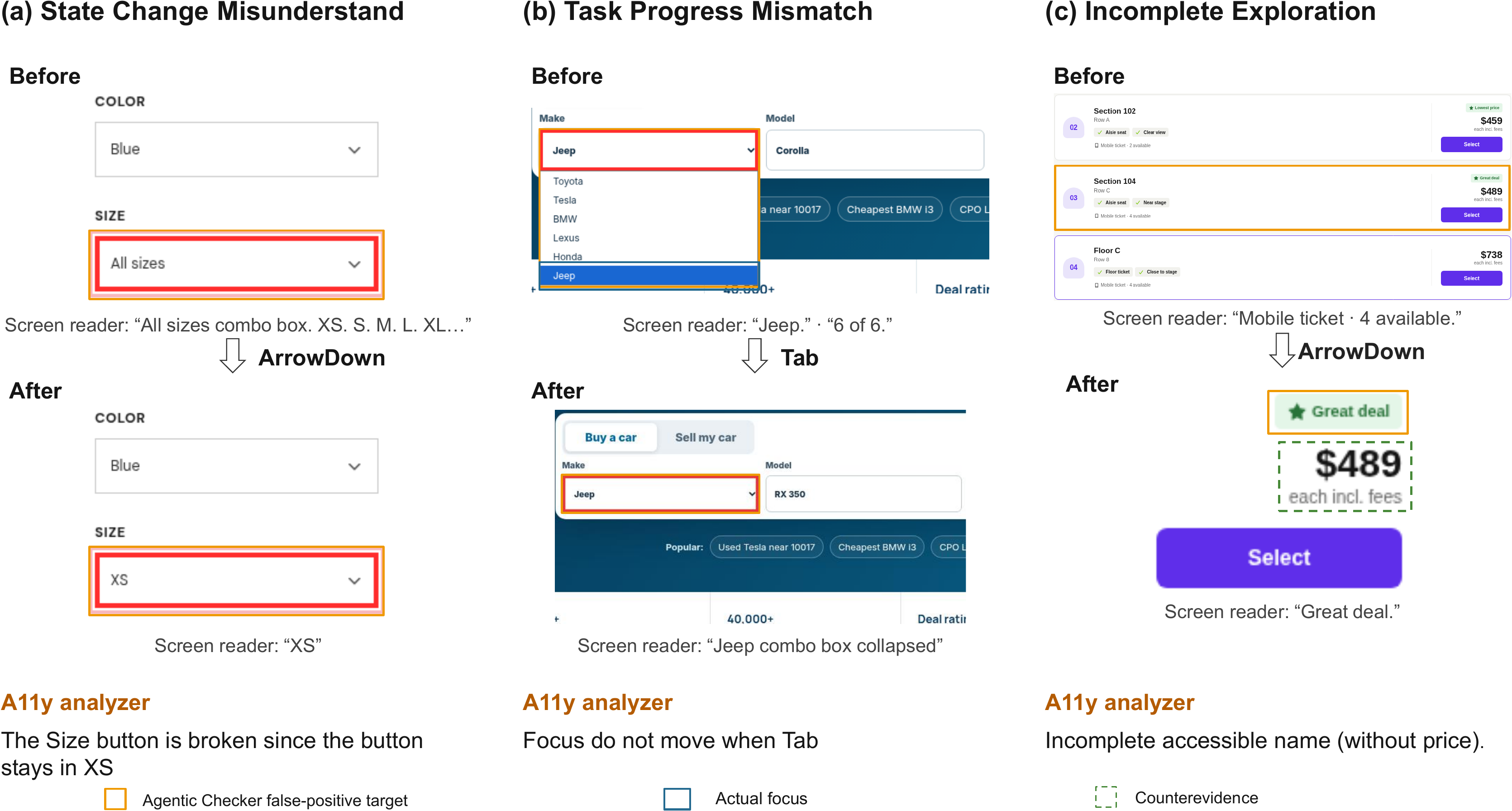}
    \caption{Representative false positives from the agentic checker: 
(a) misunderstanding an interface state change, 
(b) misinterpreting task progress as a keyboard-focus failure, and 
(c) making a judgment from incomplete screen-reader exploration. 
Orange boxes indicate the evidence targeted by the analyzer's report; green dashed boxes highlight counterevidence identified during manual review.}
\Description{Three panels illustrate false positives caused by interpreting keyboard and screen-reader traces. Panel a shows ArrowDown changing a size selection from All sizes to XS; the analyzer incorrectly treats XS as the starting value and reports no movement to S. Panel b shows Tab closing an expanded vehicle Make selector; the analyzer interprets this intermediate state as a focus trap preventing access to Model. Panel c shows a ticket listing and the utterance Great deal; the analyzer concludes that accessible ticket information is missing, while a dashed box highlights the price of 489 dollars each, including fees. The accompanying text explains that subsequent navigation reaches Model and that further screen-reader output supplies the ticket price. Solid boxes mark evidence targeted by the analyzer and dashed boxes mark counterevidence.}
    \label{fig:taskaudit-false-positives}
\end{figure}

Figure \ref{fig:taskaudit-false-positives} illustrates three representative false-positive cases from the agentic checker and highlights how errors can arise when the accessibility analyzer interprets traces of task executor. Across these examples, the analyzer either misinterprets intermediate keyboard state changes, treats incomplete screen-reader observations as evidence of missing information, or reaches a judgment that is inconsistent with the progress already made during exploration.

\paragraph{Misunderstanding interface state changes.}
The accessibility analyzer can infer the wrong interface state from screen-reader output. In the size-selector example, the selected value before the action is ``All sizes,'' while the screen reader also announces available options including ``XS,'' ``S,'' and ``M.'' After ArrowDown, the selected value changes to ``XS.'' The analyzer treats ``XS'' from the earlier screen-reader output as the pre-action value and therefore concludes that the selection did not change. The false positive results from misunderstanding which part of the screen-reader output represents the current interface state.

\paragraph{Mismatching task progress with focus movement.}
The analyzer can also interpret task progress differently from the executor. In the vehicle-search example, the Make combobox is expanded with ``Jeep'' selected before Tab. After Tab, the selection is retained and the combobox collapses, which the screen reader announces as ``Jeep combo box collapsed.'' The analyzer nevertheless reports that focus did not move when Tab was pressed. Here, it treats the absence of immediate focus relocation as a failure without accounting for the interaction progress made by closing the expanded combobox and completing the current selection.

\paragraph{Judging from incomplete exploration.}
The analyzer may make an accessibility judgment before the executor has explored enough of the interface. In the ticket-listing example, the screen reader first announces ``Mobile ticket, 4 available'' and then, after ArrowDown, ``Great deal.'' The analyzer concludes from this partial observation that the accessible name is incomplete because it does not include the price. However, the ticket also exposes the price and fee information as the user continues through the listing. The false positive therefore arises from treating information not yet encountered during sequential screen-reader exploration as information that is absent from the interface.

Together, these cases show that the agentic checker's false positives often arise from uncertainty in interpreting execution traces. Misidentified interface states, mismatched interpretations of interaction progress, and incomplete screen-reader exploration can cause successful interactions or available information to be judged as accessibility failures.

\section{Discussion}

Our results demonstrate that using temporal property checkers in websites is effective for automatically detecting navigation errors, with A11yLTLNav achieving 88.7\% precision across websites. The key insight behind A11yLTLNav is to evaluate accessibility not only from individual interface states, but also from how those states change in response to user interactions. By combining random exploration with temporal properties derived from empirical accessibility failures, A11yLTLNav can systematically check interaction expectations involving focus, keyboard interaction, widget state, and user feedback as they arise during exploration. In this section, we discuss how temporal property testing complements existing accessibility evaluation methods, how empirical accessibility findings can inform executable accessibility checks, and the limitations and opportunities for extending this approach.

\subsection{Complementing Existing Accessibility Checkers}

Our results suggest that temporal property testing complements existing approaches to automate accessibility checking. Rule-based checkers efficiently identify many violations that can be determined from a single interface state, but automated checking covers only a subset of accessibility failures~\cite{vigo2013benchmarking,fischer2025coverage}. In contrast, many failures targeted by A11yLTLNav arise from how an interface changes after an interaction, consistent with prior work on automatically detecting keyboard and dialog related navigation failures~\cite{chiou2023bagel,chiou2023detecting}. For example, opening a dialog requires relating its appearance to the resulting focus state, while activating a control may require determining whether an accessibility-relevant change or feedback follows. Temporal properties make these action--state relationships explicit and allow interaction expectations to be checked directly.

Our comparison with the agentic checker further highlights complementary forms of accessibility evidence. A11yLTLNav is well suited to failures whose expected state transitions can be specified explicitly, whereas agentic checkers can incorporate screen-reader output and task context to reason about outcomes that depend on broader semantic interpretation~\cite{zhong2025screenaudit,zhong2026taskaudit}. These approaches therefore occupy different roles in accessibility checking process: rule-based tools check local conformance, temporal properties check interaction transitions, and agentic checker can reason about richer task-specific and user-perceived outcomes. Combining these forms of evidence may enable automated evaluation to cover a broader range of accessibility failures than any single approach alone.

\subsection{Limitations and Future Work}

\paragraph{Observable evidence and assistive-technology feedback.}
A11yLTLNav currently focuses on accessibility failures whose expected behavior can be grounded in browser-observable interaction states. This scope allows us to check failures involving focus, keyboard operation, widget state, and accessibility-relevant feedback, but it does not cover all failures in our taxonomy. Some failures require evidence outside the current state model, such as actual screen-reader output, assistive-technology behavior, semantic relationships between interface regions, or interaction state that persists across views. For example, A11yLTLNav can determine whether an interaction produces a feedback signal that can be exposed to screen reader, but it does not verify what a screen reader actually announces. Future work could extend the state extractor with screen-reader and other assistive-technology observations, richer semantic relationships between interface elements, and persistent interaction state. These additional observations could allow more empirically observed accessibility failures in our taxonomy to be expressed as executable temporal properties.

\paragraph{Task-dependent accessibility properties.}
The current temporal properties are largely reusable and task-independent. This makes the same property applicable across websites, but limits A11yLTLNav's ability to check failures whose expected behavior depends on the user's goal, such as whether a description provides task-relevant information, whether controls occur in an appropriate task order, or whether a navigation path is sufficiently efficient. Agentic accessibility checkers can incorporate richer task and screen-reader context when reasoning about such outcomes~\cite{zhong2025screenaudit,zhong2026taskaudit}, suggesting a complementary direction for temporal checking. Future systems could provide parameterized temporal-property templates whose task-specific predicates are instantiated during agent execution. An agent could identify the task-relevant target, expected outcome, or relationship between interface regions, while the temporal property provides an explicit condition for checking the resulting interaction. This could combine the semantic flexibility of task-driven agents with the explicit and reusable checking boundaries of temporal properties.

\paragraph{Coverage under stochastic exploration.}
Because A11yLTLNav relies on stochastic exploration, an implemented property can be evaluated only when the explorer reaches an interaction that triggers it. Consequently, failing to report a violation within a fixed exploration budget does not establish that the violation is absent, and different exploration runs may exercise different portions of the interface. Our evaluation therefore reports the precision and stability of reported findings rather than exhaustive coverage of all failures on a website. Future work could make exploration property-aware by prioritizing states and action sequences that exercise properties that have not yet been triggered. Such guidance could be particularly useful for properties that require multiple related interactions, such as opening a dialog, navigating within it, and dismissing it to evaluate focus return. Property-guided exploration could therefore improve the coverage of executable properties within a fixed testing budget.

\paragraph{From detection to repair.}
This work focuses on detecting accessibility failures and does not evaluate whether detected violations can be repaired safely. This distinction is increasingly relevant as GenAI-based authoring tools like coding agents introduce new accessibility workflows for BLV and keyboard-only users~\cite{oswal2024examining}, while recent LLM-based accessibility repair systems show that generated fixes require explicit validation to preserve interface behavior and avoid regressions~\cite{mowar2025codea11y, wanscher2026blind,DBLP:journals/corr/abs-2605-27716}. Prior accessibility checkers have also demonstrated the value of integrating concrete accessibility improvements into developer workflows and supporting collaboration around proposed changes~\cite{DBLP:conf/w4a/BighamL07}. Building on these directions, A11yLTLNav could serve as a continuous feedback mechanism alongside a generative UI system or coding agent: after a violation is detected, the system could return its triggering action, observed states, and expected transition, and then re-check a proposed revision before it is accepted. This would form a loop, generation-execution-checking-repair, in which generated fixes are validated against explicit behavioral expectations rather than accepted solely from model-generated reasoning. Such workflows should also ensure that the authoring and feedback interfaces themselves remain accessible to BLV and keyboard-only users~\cite{oswal2024examining}.

\section{Conclusion}

In this work, we introduced a failure-driven, property-based approach to automatically detecting accessibility navigation errors experienced by blind and low-vision users using screen readers when navigating the websites. We first organized empirically observed Web accessibility failures into a taxonomy and translated the browser-observable subset into executable temporal properties. Building on these properties, A11yLTLNav combines state-aware keyboard exploration with runtime property monitoring to detect accessibility failures that emerge across interaction transitions. Across 31 website environments, A11yLTLNav achieved 88.7\% precision and remained stable across three independent exploration runs, while automating the checking process of some kinds of failures that complementary to both rule-based and agentic accessibility checkers. These results demonstrate that accessibility navigation failures involving focus, interface state, and accessible feedback can be expressed as reusable temporal properties. More broadly, our work provides a path for translating empirical knowledge from BLV accessibility research into executable properties, complementing existing accessibility evaluation methods and providing a foundation for future property-guided testing and generation--checking--repair workflows.

\begin{acks}

\end{acks}


\bibliographystyle{ACM-Reference-Format}
\bibliography{references}


\appendix
\section{LTL Semantics}
\label{appendix::ltl}

\newcommand{\traceref}[2]{#1_{#2}}
\newcommand{\tracesuffix}[2]{#1\mathrel{>\!\!>}#2}
\newcommand{\sembox}[1]{\makebox[2cm][l]{#1}}
\newcommand{\code}[1]{\texttt{#1}}

Linear Temporal Logic (LTL) specifies the behavior of systems that change over time.
The LTL syntax that we use in this paper extends propositional logic with four temporal operators.
A propositional formula $P$ is one of the following:
    an atom \code{x},
    a conjunction $P_0~\wedge~P_1$,
    a disjunction $P_0~\vee~P_1$,
    an implication $P_0~\rightarrow~P_1$, or
    a negation $\neg~P_0$.
The atoms come from some finite set; typically, one that describes a dynamic system.
An LTL formula $Q$ is one of the following:

\begin{itemize}
  \item
    a propositional formula $P$, or a conjunction, disjunction, implication, or negation of LTL formulas;
  \item
    an ``always'' formula $\code{G(}Q_0\code{)}$;
  \item
    an ``eventually'' (or ``finally'') formula $\code{F(}Q_0\code{)}$;
  \item
    a ``next'' formula $\code{X(}Q_0\code{)}$; or
  \item
    an ``until'' formula $Q_0~\code{U}~Q_1$.
\end{itemize}

Semantically, an LTL formula describes a set of \emph{traces}, each of which describes one run of the system under test.
We model traces as infinite sequences of sets of atoms.
For example, the trace $T = \traceref{T}{0}, \traceref{T}{1}, \traceref{T}{2}, \ldots$ where $\traceref{T}{i} = \{\code{Red}\}$ if $i$ is even and $\traceref{T}{i} = \{\code{Green}\}$ if $i$ is odd describes an execution of a stoplight in which the Red and Green lights alternate forever.

Let $\tracesuffix{T}{k}$ be the suffix of trace $T$ obtained by dropping the first $k$ elements.
For example, $\tracesuffix{T}{0} = T$ and $\tracesuffix{T}{2} = \traceref{T}{2}, \traceref{T}{3}, \ldots$ (dropping $\traceref{T}{0}$ and $\traceref{T}{1}$).
A temporal formula $Q$ is satisfied by a trace $T$, written $T \vDash Q$, under the following conditions:

\begin{itemize}
  \item
    \sembox{$T \vDash \code{G(}Q_0\code{)}$} if $\tracesuffix{T}{k} \vDash Q_0$ for all $k \in \mathbb{N}$;
  \item
    \sembox{$T \vDash \code{F(}Q_0\code{)}$} if $\tracesuffix{T}{k} \vDash Q_0$ for some $k \in \mathbb{N}$;
  \item
    \sembox{$T \vDash \code{X(}Q_0\code{)}$} if $\tracesuffix{T}{1} \vDash Q_0$;
  \item
    \sembox{$T \vDash Q_0~\code{U}~Q_1$} if $\tracesuffix{T}{k} \vDash Q_1$ for some $k \in \mathbb{N}$
    and $\tracesuffix{T}{j} \vDash Q_0$ for all $j < k$.
\end{itemize}

\section{PRISMA records}

ACM Digital Library search results (July 2026). IEEE Xplore search results (July 2026). Springer Nature Link.
\begin{quote}
\small\ttfamily
("blind" OR "low vision" OR "screen reader" OR "visually impaired") AND
("web" OR "webpage" OR "website" OR "browser") AND
("failure" OR "barrier" OR "error" OR "difficulty" OR "problem" OR "inaccessible" OR "usability")
\end{quote}

Special selection due to system search of Springer Nature Link. 
- Content type → **Article**
- Language → **English**
- Subjects → **Interaction design**, **Web accessibility evaluation and user experience**, **User interfaces and human computer interaction**
- Subdisciplines → User interfaces and human computer interaction

Elsevier ScienceDirect. Keywords varied since the limitations of figures via search ("Use fewer boolean connectors (max 8 per field").
\begin{quote}
\small\ttfamily
("blind" OR "visually impaired" OR "low vision") AND ("web" OR "website") AND ("accessibility" OR "barrier" OR "failure" OR "usability")
\end{quote}
- Content type → Research Articles

\begin{quote}
\small\ttfamily
Q1:
("blind" OR "low vision" OR "screen reader" OR "visually impaired")
AND ("web" OR "webpage" OR "website" OR "browser" OR "web application")
AND ("failure" OR "barrier" OR "error" OR "difficulty" OR "problem" 
     OR "inaccessible" OR "usability" OR "challenge" OR "obstacle")
\\
Q2:
("blind" OR "screen reader" OR "visually impaired")
AND ("web" OR "website" OR "browser")
AND ("keyboard" OR "navigation" OR "interaction" OR "focus")
AND ("failure" OR "barrier" OR "problem" OR "challenge")
\end{quote}

\begin{table}[h]
\centering
\small
\caption{Record counts and deduplication prior to screening. Q1 yielded 112 records (abstract field) and 3 records (title field); Q2 yielded 16 records, all subsumed by Q1. After deduplication, 113 unique records remained for screening.}
\label{tab:search-total}
\begin{tabular}{lrrr}
\toprule
\textbf{Source / step} & \textbf{$n$}  & Reduced & Screen Sampling \\
\midrule
ACM DL Abstract + Title search  & 196 &$-6$ (not papers)& 6\\
IEEE Xplore Abstract search     & 138 & $-4$(not papers) & 4 \\
Springer            & 383 & & 12\\
ScienceDirect            & 238 & $-2$ (Duplicate) & 8 \\
Q1+Q2       & 115+16 & $-18$ (Duplicate) & 0 \\

\midrule
\textbf{Total records for screening}   & & \textbf{1,056} & 30\\
\bottomrule
\end{tabular}\end{table}

\section{A11y-CUA Trace Annotation}
\label{appendix:a11y-cua-annotation}

While our systematic literature review captures a broad foundation of accessibility barriers, published studies often report broad findings that lack the step-by-step execution traces needed for formalization~\cite{huq2026bridging}. Furthermore, taxonomies from previous literature may not fully reflect the temporal navigation errors that appear in highly dynamic websites~\cite{fernandes2012evaluating}. To complement our literature synthesis with concrete evidence, we analyzed empirical interaction traces from the A11y-CUA dataset~\cite{a11y-cua}, which records blind and low-vision (BLV) users completing 12 everyday tasks on websites using screen readers and keyboard navigation.

Rather than treating task completion as a single binary outcome, we analyzed the traces as sequences of individual actions. Specifically, we identified cases where users' navigation was interrupted, took unnecessary steps, or led to unexpected interface states that they had to recover from~\cite{lazar2007frustrates}. For each case, we recorded the task context, the interactions leading to the navigation error, and its effect on user progress.

We then compared navigation errors across tasks and grouped cases with similar navigation problems into general failure types. For example, repeatedly navigating through irrelevant page structures was categorized as excessive sequential navigation, while unexpected focus jumps back to earlier parts of the page were categorized as focus reset failures. This grouping captures frequent navigation problems across different websites and tasks. Finally, we focused on navigation errors caused by interface behavior, including keyboard navigation, focus management, widget interaction, and the exposure of dynamic elements, while excluding cases caused by user mistakes or operating system behavior. These failure types, derived from prior literature and real-world traces, form the basis for our temporal properties.

\section{Detailed Failure Taxonomy}
\label{sec:detailed_failure_taxonomy}
Table~\ref{tab:taxonomy} shows the full failure taxonomy from L1 to L3 levels. 

\newcolumntype{L}[1]{>{\RaggedRight\arraybackslash}p{#1}}
\begingroup
\scriptsize
\setlength{\tabcolsep}{3.5pt}
\renewcommand{\arraystretch}{1.12}
\begin{longtable}{@{}L{2.00cm} L{4.45cm} L{5.70cm} L{1.90cm}@{}}
\caption{Taxonomy of BLV web accessibility failures, organised by the level of evidence required to
observe them (Table~\ref{tab:eval-levels}). Levels are cumulative rather than compositional: an L3
failure persists even when every L1 and L2 check passes. A failure domain therefore recurs across
levels---a missing text alternative is an L1 failure, a state change that is never announced is an L2
failure, and a description that is present yet does not reveal goal-relevant information is an L3
failure. \textit{Evidence} cites the reporting studies, including our own A11y-CUA sessions with
blind screen-reader participants~\cite{a11y-cua}. Appendix~\ref{appendix:a11y-cua-annotation} describes our trace annotation procedure for A11y-CUA sessions. The temporal properties instantiated for the
formalised subset are given in Table~\ref{tab:properties}.}
\label{tab:taxonomy_appendix}\\
\toprule
\textbf{Category} & \textbf{Failure} & \textbf{Example} & \textbf{Evidence}\\
\midrule
\endfirsthead
\multicolumn{4}{@{}l}{\scriptsize\itshape Table~\ref{tab:taxonomy}, continued}\\[1pt]
\toprule
\textbf{Category} & \textbf{Failure} & \textbf{Example} & \textbf{Evidence}\\
\midrule
\endhead
\midrule
\multicolumn{4}{r@{}}{\scriptsize\itshape continued on the next page}\\
\endfoot
\bottomrule
\endlastfoot
\rowcolor{black!88}\multicolumn{4}{@{}l@{}}{\textcolor{white}{\rule{0pt}{2.1ex}\textbf{L1: Element-level}\quad\textit{decidable by rule-based checking of the rendered page}}}\\[1pt]
\texttt{Text alternatives} & No alt text, label, or accessible name & an image, chart or animation with no \texttt{alt} attribute; a form field whose visible label is not bound to the control; a link whose only content is an icon; a CAPTCHA offering no audio route & \cite{tomlinson2016perceptions,yu2011us,zhang2024enhancing,fakrudeen2025evaluation,csontos2021accessibility,yu2025cluttered,leporini2011google,power2012guidelines,brady2013investigating,perera2026m,uckun2020breaking,buzzi2010facebook,harper2000pilot,ryskeldiev2022investigating,islam2010mixture,loiacono2009state,marques2023description,zhang2022ga11y, loiacono2009state, gleason2020twitter, wentz2011usability, jeong2023wataa, de2013web, bigham2006webinsight, asakawa2005s, whitelaw2003make}\\
\addlinespace[1.5pt]
\texttt{Structure and semantics} & No headings or landmarks & a page built entirely from \texttt{<div>} and \texttt{<span>}, offering nothing to jump between & \cite{buzzi2010facebook,brudvik2008hunting,takagi2007analysis,yu2025cluttered,a11y-cua,whitelaw2003make}\\
 & Improper heading levels or structural tags & a section heading marked \texttt{<h4>} purely because that size looked right; every heading on the page at the same level & \cite{leporini2011google,yu2025cluttered,gadde2014screen}\\
 & No skip link past the navigation & the same forty navigation links must be traversed on every page before the content begins & \cite{pakdeechote2012new,a11y-cua,ryskeldiev2022investigating}\\
 & Broken or duplicated links & a product thumbnail and its title announced as two separate links to the same page; a menu entry leading to a dead URL & \cite{csontos2021accessibility,power2012guidelines, asakawa2005s, asakawa2005s,whitelaw2003make,pascual2014impact}\\
\addlinespace[1.5pt]
\texttt{Keyboard} & No keyboard path to a control & a \texttt{<div>} styled as a button that Tab never reaches; a slider that responds only to dragging & \cite{fakrudeen2025evaluation,oswal2024examining,pascual2014impact}\\
\addlinespace[2pt]
\rowcolor{black!88}\multicolumn{4}{@{}l@{}}{\textcolor{white}{\rule{0pt}{2.1ex}\textbf{L2: Interaction-level}\quad\textit{surfaces only when the page is driven; invisible to static checking}}}\\[1pt]
\texttt{Keyboard and focus} & Loss of focus after activation & after confirming a departure airport, focus returns to the top of the page and the date fields must be found again & \cite{a11y-cua,harper2003middleware,fakrudeen2025evaluation,pakdeechote2012new,harper2000pilot,huq2026bridging}\\
 & No exit from a composite widget & Tab is intercepted inside a video description panel and never advances to the next control & \cite{a11y-cua}\\
 & Improperly hidden elements remain focusable & Tab stops on items in a collapsed carousel and the screen reader announces an empty string & \cite{a11y-cua, wentz2011usability,whitelaw2003make}\\
\addlinespace[1.5pt]
\texttt{Composite widgets} & No valid active option in an expanded combobox & the suggestion list is open but no option is marked as current, so nothing is announced while arrowing & \cite{a11y-cua, wentz2011usability}\\
 & No active-option movement on arrow keys & pressing Down inside an open suggestion list does not advance to the next entry & \cite{a11y-cua,whitelaw2003make}\\
 & No commit state after option selection & Enter on a highlighted airport leaves the field showing a partial string, and the user believes the choice was taken & \cite{a11y-cua, wentz2011usability}\\
 & No valid active cell in a managed grid & a date picker opens with no cell marked as current, so arrow keys have no starting point & \cite{a11y-cua}\\
 & No active-cell movement on arrow keys & Tab passes over the whole calendar, and the arrow keys that would move within it are never announced & \cite{a11y-cua}\\
 & No effect from a role-specific activation key & a calendar cell is selected only by Space; pressing Enter repeatedly produces nothing & \cite{a11y-cua,pascual2014impact}\\
\addlinespace[1.5pt]
\texttt{Feedback and state} & No feedback confirming an activation & after submitting a form nothing is announced, and the user re-reads the page to find out whether it was accepted & \cite{buzzi2010facebook,leporini2011google,huq2026bridging,power2012guidelines,ryskeldiev2022investigating, wentz2011usability,borodin2008s}\\
 & Misplaced or disconnected response to an activation & a filter updates a results panel further down the page with nothing tying the two together; correcting a typo re-opens the suggestion list on every keystroke & \cite{sandnes2024consent,harper2000pilot,uckun2020breaking,borodin2010more,a11y-cua, wentz2011usability,borodin2008s,whitelaw2003make}\\
 & ARIA state contradicts the native state & a checkbox reported as \texttt{aria-checked="false"} while the native control is checked & \cite{a11y-cua}\\
 & No exposed expanded state after a reveal & an accordion panel opens visually but the trigger continues to report itself as collapsed & \cite{a11y-cua, asakawa2005s}\\
 & No trace of operation progress or mode & an upload gives no sign of whether it is running or has stalled; the same key both inspects and edits a block, with nothing saying which just happened & \cite{huq2026bridging,li2026content,wentz2011usability,wentz2011usability}\\
\addlinespace[1.5pt]
\texttt{Dynamic content} & No accessible name or named dismissal control on a dialog & a consent banner is announced only as \emph{dialog}, with no way to tell what it is or how to close it & \cite{gupta2005extracting,wentz2011separate,leporini2011google,csontos2021accessibility,a11y-cua,borodin2010more}\\
 & No focus move into a newly opened dialog & a cookie notice appears over the page while focus stays behind it, so the user keeps reading content that is now blocked & \cite{a11y-cua,borodin2010more, wentz2011usability}\\
 & No dialog dismissal or focus return on Escape & Escape does not close the overlay; when it finally closes, focus is left at the top rather than on the control that opened it & \cite{a11y-cua}\\
 & No reachable control to start playing or stop playing media & a video begins on load and its pause control is neither reachable by Tab nor announced & \cite{tomlinson2016perceptions,miyashita2007making,pascual2014impact}\\
\addlinespace[1.5pt]
\texttt{Structure and semantics} & Inadequate labels to reveal nearby context & a price is read out with no indication of which product row or which column it belongs to & \cite{buzzi2010facebook, wentz2011usability,borodin2008s,asakawa2005s,whitelaw2003make,leuthold2008beyond}\\
\addlinespace[1.5pt]
\texttt{AT interoperability} & Inconsistent name exposure across reading modes & a password field is announced by name while browsing, but only as \emph{edit box} once the user starts typing & \cite{buzzi2010facebook,potluri2021examining}\\
 & Screen-reader shortcut captured outside the page & the context-menu command opens the browser's own menu instead of the page's; an injected accessibility widget claims keys the user's screen reader needs & \cite{a11y-cua,makati2024promise}\\
\addlinespace[2pt]
\rowcolor{black!88}\multicolumn{4}{@{}l@{}}{\textcolor{white}{\rule{0pt}{2.1ex}\textbf{L3: Task-level}\quad\textit{persists when L1 and L2 checks pass; the failure lies in the fit between structure and goal}}}\\[1pt]
\texttt{Feedback and state} & No guidance for recovering from a rejected action & a submission is refused with no indication of which field was wrong or what format it expected; a category must exist before an expense can be added, and nothing says so & \cite{huq2026bridging,ryskeldiev2022investigating,borodin2008s}\\
\addlinespace[1.5pt]
\texttt{Text alternatives} & Inadequate description of goal-relevant information & a binary tree described as running prose, leaving parent and child relations unrecoverable; a chart summarised in one sentence when the task requires comparing series; a link named \emph{read more} among twenty others & \cite{ferreira2012aligning,power2012guidelines,yu2025cluttered,fukuda2005proposing,harper2003middleware,alam2023seechart,loiacono2009state,l2026nonvisual,yesilada2003rendering,krejtz2025higher,uckun2020breaking,ryskeldiev2022investigating,islam2010mixture,borodin2010more,leporini2011google,kodandaram2026finding, loiacono2009state, gleason2020twitter, jeong2023wataa, bigham2006webinsight, whitelaw2003make, pascual2014impact,leuthold2008beyond}\\
\addlinespace[1.5pt]
\texttt{Structure and semantics} & Reading order and grouping inconsistent with the task's logic & a product's price is read far from its name; the store-locator sits under \emph{Account} rather than under \emph{Stores} & \cite{yu2025cluttered,gadde2014screen,potluri2021examining,leporini2011google,ryskeldiev2022investigating,krejtz2025higher,pakdeechote2012new,power2012guidelines,ferreira2012aligning,aizpurua2015prejudices,wentz2011separate,yang2013bypassing}\\
 & Excessive heading marks relative to page content & every list item carries a heading, so the heading list is as long as the page & \cite{yu2025cluttered,gadde2014screen}\\
 & Tab order inconsistent with the task's order & the submit button is reached before the fields it depends on & \cite{fakrudeen2025evaluation, asakawa2005s}\\
 & Position learned from prior visits no longer predicts where the control is & the submit button sits in the top bar on one page and after the last field on the next & \cite{huq2026bridging, wentz2011usability}\\
\addlinespace[1.5pt]
\texttt{Navigation efficiency} & No efficient route to the control the task requires & reaching a store-selection setting takes several hundred Tab presses through the footer & \cite{fukuda2005proposing,ferdous2023enabling,power2012guidelines,da2018content,potluri2021examining,a11y-cua,ryskeldiev2022investigating,borodin2010more, wentz2011usability}\\
 & No confirmation of completeness, driving re-verification & a forum thread is re-read from the top because nothing confirms that no later reply revised the answer & \cite{gadde2014screen,kodandaram2026finding,harper2003middleware,borodin2010more,bigham2017effects,borodin2008s}\\
 & No means of resuming from a prior position & returning from a product page drops the user at the top of the results list rather than at the item they left & \cite{rohani2012back,prakash2023autodesc,borodin2008s}\\
\end{longtable}
\endgroup
\section{Failure-to-Temporal-Property Mapping}
\label{appendix:failure-property-mapping}

For readability, we assign compact property identifiers by taxonomy level, like \texttt{L1-01}, \texttt{L2-01}, and so on. Table~\ref{tab:failure-property-map}
shows how the failure categories formalized by A11yLTLNav map to executable
temporal properties. A failure may map to more than one property when distinct
observable interaction mechanisms require separate checks. Failure categories
that are outside the current executable property set remain in the full taxonomy
appendix and are not repeated here.

\begingroup
\footnotesize
\setlength{\tabcolsep}{3.5pt}
\renewcommand{\arraystretch}{1.15}
\begin{longtable}{@{}>{\RaggedRight\arraybackslash}p{4.2cm}
                        >{\centering\arraybackslash}p{1.3cm}
                        >{\RaggedRight\arraybackslash}p{8.6cm}@{}}
\caption{Mapping from formalized accessibility failures to simplified temporal
properties. Each row represents one executable property; a blank \emph{Failure}
cell continues the failure immediately above.}
\label{tab:properties}\label{tab:failure-property-map}\\
\toprule
\textbf{Failure} & \textbf{ID} & \textbf{Simplified temporal property}\\
\midrule
\endfirsthead

\multicolumn{3}{@{}l}{\footnotesize\itshape Table~\ref{tab:failure-property-map}, continued}\\[1pt]
\toprule
\textbf{Failure} & \textbf{ID} & \textbf{ Temporal property}\\
\midrule
\endhead

\midrule
\multicolumn{3}{r@{}}{\footnotesize\itshape continued on the next page}\\
\endfoot

\bottomrule
\endlastfoot

\rowcolor{black!88}\multicolumn{3}{@{}l@{}}{\textcolor{white}{\rule{0pt}{2.1ex}\textbf{L1: Element-level}}}\\[1pt]
No alt text, label, or accessible name
& \texttt{L1-01}
& \(\mathbf{G}(\mathrm{exposed}(x)\land\mathrm{nameReq}(x)\rightarrow\mathrm{hasNameOrTextAlt}(x))\)\\
\addlinespace[1.2pt]
No headings or landmarks
& \texttt{L1-02}
& \(\mathbf{G}(\mathrm{documentComplete}\rightarrow\mathrm{orientationAnchorCount}>0)\)\\
\addlinespace[1.2pt]
Improper heading levels or structural tags
& \texttt{L1-03}
& \(\mathbf{G}(\mathrm{documentComplete}\rightarrow\neg\mathrm{forwardHeadingSkip})\)\\
\addlinespace[1.2pt]
No skip link past the navigation
& \texttt{L1-04}
& \(\mathbf{G}(\mathrm{longPreMainPrefix}\rightarrow\mathrm{bypassPresent})\)\\
\addlinespace[1.2pt]

& \texttt{L1-05}
& \(\mathbf{G}(\mathrm{activateBypass}(c,t)\rightarrow\mathbf{X}\,\mathrm{validContinuation}(t))\)\\
\addlinespace[1.2pt]
Broken or duplicated links
& \texttt{L1-06}
& \(\mathbf{G}(\mathrm{documentComplete}\rightarrow\neg\mathrm{redundantLocalLinkPair})\)\\
\addlinespace[1.2pt]
No keyboard path to a control
& \texttt{L1-07}
& \(\mathbf{G}(\mathrm{eligibleControl}(c)\rightarrow\mathrm{tabReachable}(c))\)\\
\addlinespace[1.2pt]
\addlinespace[2pt]
\rowcolor{black!88}\multicolumn{3}{@{}l@{}}{\textcolor{white}{\rule{0pt}{2.1ex}\textbf{L2: Interaction-level}}}\\[1pt]
Loss of focus after activation
& \texttt{L2-01}
& \(\mathbf{G}(\mathrm{localInvalidation}(c)\rightarrow\mathbf{X}\,\mathrm{exposedFocus})\)\\
\addlinespace[1.2pt]

& \texttt{L2-02}
& \(\mathbf{G}(\mathrm{majorViewTransition}(c)\rightarrow\mathbf{X}\,\mathrm{orientationCue}(c))\)\\
\addlinespace[1.2pt]
No exit from a composite widget
& \texttt{L2-03}
& \(\mathbf{G}(\mathrm{blockedTab}(c)\rightarrow\mathbf{X}\,\mathrm{focusMoves}(c))\)\\
\addlinespace[1.2pt]
Improperly hidden elements remain focusable
& \texttt{L2-04}
& \(\mathbf{G}(\mathrm{trustedTab}\rightarrow\mathbf{X}(\neg\mathrm{focusHidden}))\)\\
\addlinespace[1.2pt]
No valid active option in an expanded combobox
& \texttt{L2-05}
& \(\mathbf{G}(\mathrm{expandedCombobox}(c)\rightarrow\mathrm{validActiveOption}(c))\)\\
\addlinespace[1.2pt]
No active-option movement on arrow keys
& \texttt{L2-06}
& \(\mathbf{G}(\mathrm{comboboxArrow}(c,k,n)\rightarrow\mathbf{X}(\mathrm{activeDescendant}(c)=n))\)\\
\addlinespace[1.2pt]
No commit state after option selection
& \texttt{L2-07}
& \(\mathbf{G}(\mathrm{enterOnActiveOption}(c,o)\rightarrow\mathbf{X}(\mathrm{committed}(c,o)))\)\\
\addlinespace[1.2pt]
No valid active cell in a managed grid
& \texttt{L2-08}
& \(\mathbf{G}(\mathrm{managedGrid}(g)\rightarrow\mathrm{validActiveCell}(g))\)\\
\addlinespace[1.2pt]
No active-cell movement on arrow keys
& \texttt{L2-09}
& \(\mathbf{G}(\mathrm{gridArrow}(g,k,n)\rightarrow\mathbf{X}(\mathrm{activeCell}(g)=n))\)\\
\addlinespace[1.2pt]
No effect from a role-specific activation key
& \texttt{L2-10}
& \(\mathbf{G}(\mathrm{roleActivation}(c,k,e)\rightarrow\mathbf{X}(\mathrm{exactEffect}(e)))\)\\
\addlinespace[1.2pt]
No feedback confirming an activation
& \texttt{L2-11}
& \(\mathbf{G}(\mathrm{effectfulActivation}(c)\rightarrow\mathbf{X}\,\mathrm{notifiableFeedback}(c))\)\\
\addlinespace[1.2pt]
ARIA state contradicts the native state
& \texttt{L2-12}
& \(\mathbf{G}(\mathrm{nativeCheckableARIA}(c)\rightarrow\mathrm{ariaChecked}(c)=\mathrm{nativeChecked}(c))\)\\
\addlinespace[1.2pt]

& \texttt{L2-13}
& \(\mathbf{G}(\mathrm{nativeOptionWithARIA}(o)\rightarrow\mathrm{ariaSelected}(o)=\mathrm{nativeSelected}(o))\)\\
\addlinespace[1.2pt]

& \texttt{L2-14}
& \(\mathbf{G}(\mathrm{nativeRangeWithARIA}(q)\rightarrow\mathrm{ariaValueNow}(q)=\mathrm{nativeValue}(q))\)\\
\addlinespace[1.2pt]
No exposed expanded state after a reveal
& \texttt{L2-15}
& \(\mathbf{G}(\mathrm{expansionAction}(c,r)\rightarrow\mathbf{X}(\mathrm{ariaExpanded}(c)=\mathrm{exposed}(r)))\)\\
\addlinespace[1.2pt]

& \texttt{L2-16}
& \(\mathbf{G}(\mathrm{disclosureReveal}(c,r)\rightarrow\mathbf{X}(\mathrm{ariaExpanded}(c)=\mathrm{exposed}(r)))\)\\
\addlinespace[1.2pt]
No accessible name or named dismissal control on a dialog
& \texttt{L2-17}
& \(\mathbf{G}(\mathrm{visibleDialog}(d)\rightarrow\mathrm{hasProgrammaticName}(d))\)\\
\addlinespace[1.2pt]

& \texttt{L2-18}
& \(\mathbf{G}(\mathrm{visibleDialog}(d)\rightarrow\mathrm{namedFocusableAction}(d))\)\\
\addlinespace[1.2pt]
No focus move into a newly opened dialog
& \texttt{L2-19}
& \(\mathbf{G}(\mathrm{openNewModal}(d)\rightarrow\mathbf{X}(\mathrm{focusInside}(d)))\)\\
\addlinespace[1.2pt]
No dialog dismissal or focus return on Escape
& \texttt{L2-20}
& \(\mathbf{G}(\mathrm{escapeInDialog}(d)\rightarrow\mathbf{X}(\neg\mathrm{exposed}(d)))\)\\
\addlinespace[1.2pt]

& \texttt{L2-21}
& \(\mathbf{G}(\mathrm{escapeClosesModal}(d,v)\rightarrow\mathbf{X}(\mathrm{focus}=v))\)\\
\addlinespace[1.2pt]
\addlinespace[2pt]
\rowcolor{black!88}\multicolumn{3}{@{}l@{}}{\textcolor{white}{\rule{0pt}{2.1ex}\textbf{L3: Task-level}}}\\[1pt]
No guidance for recovering from a rejected action
& \texttt{L3-01}
& \(\mathbf{G}(\mathrm{nativeRejection}(f)\rightarrow\mathbf{X}\,\mathrm{actionableGuidance}(f))\)\\
\addlinespace[1.2pt]

& \texttt{L3-02}
& \(\mathbf{G}(\mathrm{newARIAInvalid}(i)\rightarrow\mathbf{X}(\mathrm{linkedGuidance}(i)))\)\\
\addlinespace[1.2pt]
\end{longtable}
\endgroup

\subsection{Meaning and scope of the executable properties}
\label{appendix:property-meaning}

Table~\ref{tab:property-meaning} explains the operational meaning of each
property without repeating its formula. The \emph{Scope} column records only
important boundaries on what the property claims to detect; an em dash means
that no additional boundary is needed beyond the triggering conditions stated
in the formula.

\begingroup
\footnotesize
\setlength{\tabcolsep}{3.5pt}
\renewcommand{\arraystretch}{1.15}
\begin{longtable}{@{}>{\centering\arraybackslash}p{1.4cm}
                        >{\RaggedRight\arraybackslash}p{7.5cm}
                        >{\RaggedRight\arraybackslash}p{5.2cm}@{}}
\caption{Operational meaning and scope of the temporal properties in
Table~\ref{tab:failure-property-map}.}
\label{tab:property-meaning}\\
\toprule
\textbf{ID} & \textbf{What it checks} & \textbf{Scope}\\
\midrule
\endfirsthead

\multicolumn{3}{@{}l}{\footnotesize\itshape Table~\ref{tab:property-meaning}, continued}\\[1pt]
\toprule
\textbf{ID} & \textbf{What it checks} & \textbf{Scope}\\
\midrule
\endhead

\midrule
\multicolumn{3}{r@{}}{\footnotesize\itshape continued on the next page}\\
\endfoot

\bottomrule
\endlastfoot

\rowcolor{black!88}\multicolumn{3}{@{}l@{}}{\textcolor{white}{\rule{0pt}{2.1ex}\textbf{L1: Element-level}}}\\[1pt]
\texttt{L1-01}
& Every exposed element that requires a name or text alternative must expose one.
& Covers supported name-required elements; adequacy of an existing name is outside this property.\\
\addlinespace[1.2pt]
\texttt{L1-02}
& A completed content-bearing page should expose at least one heading or landmark for nonvisual orientation.
& \textemdash\\
\addlinespace[1.2pt]
\texttt{L1-03}
& The exposed heading sequence should not jump forward by more than one hierarchy level.
& Covers heading-level discontinuities, not every kind of semantically inappropriate structural markup.\\
\addlinespace[1.2pt]
\texttt{L1-04}
& A long keyboard-navigation prefix before main content should provide a usable bypass control.
& \textemdash\\
\addlinespace[1.2pt]
\texttt{L1-05}
& Activating an exact main-content bypass should establish a valid keyboard continuation at its target.
& \(\mathrm{validContinuation}(t)\) means focus reaches the target directly or the requested next Tab reaches the expected continuation at or after it.\\
\addlinespace[1.2pt]
\texttt{L1-06}
& A local image/text pair for the same destination should be exposed as one navigation opportunity.
& Covers the duplicated-local-link subtype; unreachable or dead URLs are outside this property.\\
\addlinespace[1.2pt]
\texttt{L1-07}
& Every eligible standalone semantic control should be reachable through sequential keyboard navigation.
& \(\mathrm{eligibleControl}(c)\) abbreviates an exposed, enabled standalone semantic control; \(\mathrm{tabReachable}(c)\) means sequentially keyboard-focusable. Composite-widget descendants and controls with no declared interactive semantics are outside this property.\\
\addlinespace[1.2pt]
\addlinespace[2pt]
\rowcolor{black!88}\multicolumn{3}{@{}l@{}}{\textcolor{white}{\rule{0pt}{2.1ex}\textbf{L2: Interaction-level}}}\\[1pt]
\texttt{L2-01}
& A local activation that invalidates its control should leave settled focus on a concrete exposed element.
& \(\mathrm{localInvalidation}(c)\) denotes control invalidation within a stable local view; \(\mathrm{exposedFocus}\) requires concrete connected and programmatically exposed focus.\\
\addlinespace[1.2pt]
\texttt{L2-02}
& After a major same-document view transition, the interface should establish the user's orientation in the new context.
& \(\mathrm{orientationCue}(c)\) means focus enters the new context or the persistent invoker gains a current-location cue.\\
\addlinespace[1.2pt]
\texttt{L2-03}
& If page code blocks Tab while another usable keyboard target exists, settled focus should leave the current control.
& \(\mathrm{blockedTab}(c)\) combines author-cancelled Tab with an available alternative target; this remains a single-step no-progress check rather than complete cycle detection.\\
\addlinespace[1.2pt]
\texttt{L2-04}
& Tab navigation should not land on a programmatically hidden element.
& \textemdash\\
\addlinespace[1.2pt]
\texttt{L2-05}
& An expanded managed combobox should identify one valid active option.
& \textemdash\\
\addlinespace[1.2pt]
\texttt{L2-06}
& An admitted arrow-key action should move the combobox active option to the inferable adjacent option.
& \textemdash\\
\addlinespace[1.2pt]
\texttt{L2-07}
& Pressing Enter on the active option should commit that exact option.
& \textemdash\\
\addlinespace[1.2pt]
\texttt{L2-08}
& A managed grid should expose one valid active cell.
& \textemdash\\
\addlinespace[1.2pt]
\texttt{L2-09}
& An admitted arrow-key action should move the grid active cell to the inferable adjacent cell.
& \textemdash\\
\addlinespace[1.2pt]
\texttt{L2-10}
& When a role/key pair has a declared browser effect, activating it should produce that effect.
& Generic buttons without a declared exact effect remain outside this property.\\
\addlinespace[1.2pt]
\texttt{L2-11}
& An activation that changes the interface should produce at least one AT-surfaceable feedback signal.
& \(\mathrm{notifiableFeedback}(c)\) includes a focus move, focused-control state or name change, live notification, or same-document navigation; it does not prove actual screen-reader speech.\\
\addlinespace[1.2pt]
\texttt{L2-12}
& Redundant \texttt{aria-checked} should agree with the native checked or indeterminate state.
& \textemdash\\
\addlinespace[1.2pt]
\texttt{L2-13}
& Redundant \texttt{aria-selected} should agree with the native option selected state.
& \textemdash\\
\addlinespace[1.2pt]
\texttt{L2-14}
& Redundant \texttt{aria-valuenow} should agree with the native range value.
& \textemdash\\
\addlinespace[1.2pt]
\texttt{L2-15}
& After an expansion or collapse action, \texttt{aria-expanded} should match whether the controlled region is exposed.
& \textemdash\\
\addlinespace[1.2pt]
\texttt{L2-16}
& When a local disclosure reveals content, the revealing control should expose expanded state consistent with the region.
& \textemdash\\
\addlinespace[1.2pt]
\texttt{L2-17}
& Every visible dialog should expose a non-empty programmatic name.
& \textemdash\\
\addlinespace[1.2pt]
\texttt{L2-18}
& Every visible dialog should expose at least one named, enabled, sequentially focusable action.
& Establishes an identifiable action candidate; it does not prove that the action dismisses the dialog.\\
\addlinespace[1.2pt]
\texttt{L2-19}
& Opening a new modal should move keyboard focus into that exact modal.
& \textemdash\\
\addlinespace[1.2pt]
\texttt{L2-20}
& Pressing Escape in the active dialog should close or hide that exact dialog.
& \textemdash\\
\addlinespace[1.2pt]
\texttt{L2-21}
& When Escape closes a modal and its invoker remains eligible, focus should return to that exact invoker.
& \textemdash\\
\addlinespace[1.2pt]
\addlinespace[2pt]
\rowcolor{black!88}\multicolumn{3}{@{}l@{}}{\textcolor{white}{\rule{0pt}{2.1ex}\textbf{L3: Task-level}}}\\[1pt]
\texttt{L3-01}
& When native form validation rejects submission, the form should expose actionable guidance for at least one invalid field.
& \(\mathrm{actionableGuidance}(f)\) summarizes linked accessible guidance or focused native validation guidance; arbitrary task-level prerequisites remain outside this property.\\
\addlinespace[1.2pt]
\texttt{L3-02}
& When an action newly marks a field \texttt{aria-invalid=true}, that field should expose programmatically linked guidance.
& Covers action-bound \texttt{aria-invalid} transitions, not arbitrary task-level prerequisites.\\
\addlinespace[1.2pt]
\end{longtable}
\endgroup

\subsection{Notation used in simplified temporal properties}
\label{appendix:property-notation}

Table~\ref{tab:property-notation} defines the logical symbols and variables used
in Table~\ref{tab:failure-property-map}. Domain-specific predicate names are
short descriptive atomic propositions; their operational meanings are given in
Table~\ref{tab:property-meaning}.

\begingroup
\footnotesize
\setlength{\tabcolsep}{4pt}
\renewcommand{\arraystretch}{1.12}
\begin{longtable}{@{}>{\centering\arraybackslash}p{2.0cm}
                        >{\RaggedRight\arraybackslash}p{3.8cm}
                        >{\RaggedRight\arraybackslash}p{8.3cm}@{}}
\caption{Notation used in the simplified temporal properties.}
\label{tab:property-notation}\\
\toprule
\textbf{Symbol} & \textbf{Meaning} & \textbf{Interpretation / use}\\
\midrule
\endfirsthead

\multicolumn{3}{@{}l}{\footnotesize\itshape Table~\ref{tab:property-notation}, continued}\\[1pt]
\toprule
\textbf{Symbol} & \textbf{Meaning} & \textbf{Interpretation / use}\\
\midrule
\endhead

\midrule
\multicolumn{3}{r@{}}{\footnotesize\itshape continued on the next page}\\
\endfoot

\bottomrule
\endlastfoot

\rowcolor{black!88}\multicolumn{3}{@{}l@{}}{\textcolor{white}{\rule{0pt}{2.1ex}\textbf{Temporal and logical operators}}}\\[1pt]
\(\mathbf{G}\) & Globally / always & The property must hold whenever its triggering condition occurs along the interaction trace.\\
\addlinespace[1.0pt]
\(\mathbf{X}\) & Next logical state & The next relevant post-action state after the triggering interaction.\\
\addlinespace[1.0pt]
\(\land\) & Logical AND & Both conditions must hold.\\
\addlinespace[1.0pt]
\(\lor\) & Logical OR & At least one condition must hold.\\
\addlinespace[1.0pt]
\(\neg\) & Logical NOT & The following condition must not hold.\\
\addlinespace[1.0pt]
\(\rightarrow\) & Implication & If the condition on the left occurs, the condition on the right is required.\\
\addlinespace[1.0pt]
\(=\) & Equality & Two observed values or identities must agree.\\
\addlinespace[1.0pt]
\(>\) & Greater than & Used for threshold predicates such as requiring at least one orientation anchor.\\
\addlinespace[2pt]

\rowcolor{black!88}\multicolumn{3}{@{}l@{}}{\textcolor{white}{\rule{0pt}{2.1ex}\textbf{Variables}}}\\[1pt]
\(x\) & Generic interface element & Used when a property is not tied to a particular control type.\\
\addlinespace[1.0pt]
\(c\) & Control & The interactive control that triggers or owns the relevant state.\\
\addlinespace[1.0pt]
\(k\) & Keyboard key & The key used for the interaction, such as Enter, Space, or an arrow key.\\
\addlinespace[1.0pt]
\(e\) & Expected / declared effect & The browser-observable effect associated with a role-specific activation.\\
\addlinespace[1.0pt]
\(t\) & Target & The target of a bypass or navigation control.\\
\addlinespace[1.0pt]
\(d\) & Dialog / modal & The exact dialog involved in opening, closing, naming, or focus management.\\
\addlinespace[1.0pt]
\(v\) & Invoker & The exact control that opened a modal and should receive focus again after closing.\\
\addlinespace[1.0pt]
\(i\) & Invalid field & A form field identified as invalid.\\
\addlinespace[1.0pt]
\(o\) & Option & The active or selected option in a combobox or listbox.\\
\addlinespace[1.0pt]
\(g\) & Grid & The managed grid whose active cell is being checked.\\
\addlinespace[1.0pt]
\(r\) & Controlled / revealed region & The region whose exposed state should correspond to a disclosure control.\\
\addlinespace[1.0pt]
\(q\) & Native range control & A slider or range input whose native value is compared with \texttt{aria-valuenow}.\\
\addlinespace[1.0pt]
\(n\) & Next item & The inferable adjacent option or grid cell expected after an arrow-key action.\\
\addlinespace[1.0pt]
\(f\) & Form & The native form involved in a validation rejection.\\
\addlinespace[2pt]

\rowcolor{black!88}\multicolumn{3}{@{}l@{}}{\textcolor{white}{\rule{0pt}{2.1ex}\textbf{Atomic propositions}}}\\[1pt]
Predicate names
& Descriptive atomic propositions
& Composite names abbreviate browser-observable conditions. Examples include \(\mathrm{eligibleControl}\), \(\mathrm{orientationCue}\), and \(\mathrm{notifiableFeedback}\); Table~\ref{tab:property-meaning} gives their operational meaning and scope.\\
\end{longtable}
\endgroup

\subsection{Unformalized Failure Taxonomies}
\label{appendix:unformalized-failures}

Across the 36 failure categories in our taxonomy in the main text, the current property set
operationalizes 22 with at least one executable temporal property. The remaining
14 are not omitted because they are inexpressible in LTL. Rather, an executable
property must be grounded in predicates that the checker can observe and
attribute to a concrete interaction. For these failures, at least one required
predicate depends on information that is not available from browser state alone,
such as a semantic relation between interface regions, an operation lifecycle,
assistive-technology behavior, a task or relevance oracle, a calibrated cost
threshold, or persistent cross-view identity. Table~\ref{tab:unformalized-taxonomy}
summarizes these boundaries and the additional support that would be needed to
encode them as Bombadil properties.

\begingroup
\footnotesize
\setlength{\tabcolsep}{3.2pt}
\renewcommand{\arraystretch}{1.14}
\begin{longtable}{@{}>{\centering\arraybackslash}p{1.0cm}
                        >{\RaggedRight\arraybackslash}p{4.1cm}
                        >{\RaggedRight\arraybackslash}p{4.8cm}
                        >{\RaggedRight\arraybackslash}p{4.5cm}@{}}
\caption{Taxonomy failures not directly formalized by the current browser-observable property set. These failures are not necessarily inexpressible in LTL; the limitation is that one or more predicates needed for an executable check are not currently grounded by Bombadil.}
\label{tab:unformalized-taxonomy}\\
\toprule
\textbf{Level} & \textbf{Failure} & \textbf{Why it is not directly executable} & \textbf{Additional Bombadil support needed}\\
\midrule
\endfirsthead

\multicolumn{4}{@{}l}{\footnotesize\itshape Table~\ref{tab:unformalized-taxonomy}, continued}\\[1pt]
\toprule
\textbf{Level} & \textbf{Failure} & \textbf{Why it is not directly executable} & \textbf{Additional Bombadil support needed}\\
\midrule
\endhead

\midrule
\multicolumn{4}{r@{}}{\footnotesize\itshape continued on the next page}\\
\endfoot

\bottomrule
\endlastfoot

\texttt{L2}
& Misplaced or disconnected response to an activation
& A state change can be observed, but browser state alone does not establish which region is the semantically intended response to the triggering control.
& A declared or inferred trigger-to-response-region relation that binds the activated control to the region expected to change.\\
\addlinespace[1.2pt]

\texttt{L2}
& Inadequate labels to reveal nearby context
& The checker can observe a label and surrounding elements, but it cannot infer which nearby context is necessary for the label to be meaningful.
& An element-to-context relation or semantic oracle specifying which surrounding information is required for interpretation.\\
\addlinespace[1.2pt]

\texttt{L2}
& No trace of operation progress or mode
& Checking progress or mode requires knowing when an operation starts, remains active, changes phase, and completes; those lifecycle boundaries are not generally exposed by the page.
& An operation or mode lifecycle recorder with explicit start, progress, completion, and cancellation states.\\
\addlinespace[1.2pt]

\texttt{L2}
& No reachable control to stop auto-playing media
& The checker may observe media and controls independently, but it cannot reliably determine which control stops the exact auto-playing media instance.
& A media-to-control binding plus a reachability check for the identified stop or pause control.\\
\addlinespace[1.2pt]

\texttt{L2}
& Inconsistent name exposure across reading modes
& DOM and ARIA state do not establish how a screen reader exposes the same element in browse mode versus forms or focus mode.
& Assistive-technology-level observations aligned with screen-reader mode and browser focus.\\
\addlinespace[1.2pt]

\texttt{L2}
& Screen-reader shortcut captured outside the page
& A failed shortcut may be handled by the page, browser, operating system, extension, accessibility widget, or screen reader; browser-level evidence cannot reliably attribute the capture.
& OS- and assistive-technology-level key-event tracing that identifies which component consumed the shortcut.\\
\addlinespace[1.2pt]

\texttt{L3}
& Inadequate description of goal-relevant information
& Whether an existing description is sufficient depends on what information the user's task requires, not only on whether the description is present.
& A task or relevance oracle that represents the user's information goal and judges whether the exposed description supports it.\\
\addlinespace[1.2pt]

\texttt{L3}
& Reading order and grouping inconsistent with the task's logic
& The browser exposes document order and grouping, but the checker does not know the task-dependent relationships that should determine the useful order or grouping.
& A task model that specifies relevant objects, relationships, and expected grouping or ordering.\\
\addlinespace[1.2pt]

\texttt{L3}
& Excessive heading marks relative to page content
& Heading density is observable, but there is no task-independent threshold at which the number of headings becomes excessive for navigation.
& A calibrated density or navigation-cost threshold, potentially conditioned on page structure or task context.\\
\addlinespace[1.2pt]

\texttt{L3}
& Tab order inconsistent with the task's order
& The actual Tab sequence is observable, but the checker has no general oracle for the order in which controls should be encountered for a particular goal.
& A declared task order or task-derived partial order over relevant controls.\\
\addlinespace[1.2pt]

\texttt{L3}
& No efficient route to the control the task requires
& Efficiency requires both knowing the task-relevant target and deciding how much navigation cost is acceptable; neither is defined by browser state alone.
& A task-target oracle together with a navigation-cost model or explicit cost bound.\\
\addlinespace[1.2pt]

\texttt{L3}
& No confirmation of completeness, driving re-verification
& The checker cannot infer when the user has seen enough task-relevant evidence to know that a review or search is complete.
& A task-completion model that represents relevant evidence and the conditions under which completeness can be established.\\
\addlinespace[1.2pt]

\texttt{L3}
& No means of resuming from a prior position
& Checking resumption requires a stable identity for the user's prior interaction position across navigation or re-rendering and a definition of equivalent restoration.
& Persistent cross-view position identity and history tracking that can compare the restored location with the prior task position.\\

\texttt{L3}
& Position learned from prior visits no longer predicts where the control is
& Detecting positional inconsistency requires comparing a control's location
across comparable pages or prior visits, which cannot be determined from a
single browser execution state.
& Persistent cross-view element identity together with a model of comparable
page structure or expected control location.\\
\addlinespace[1.2pt]

\end{longtable}
\endgroup

The 22 mapped failure categories should therefore be interpreted as coverage by at least one executable property, not as exhaustive coverage of every instance of that category. Several implemented properties intentionally capture narrower, high-confidence subtypes for which the trigger and expected outcome can be observed reliably in the browser. This conservative boundary keeps the properties aligned with what the Bombadil monitor can actually establish from an execution trace.

\section{Prompts for Website Generation}
\label{appendix::promptuxCUA}
We provide the complete prompt used to generate the website clones in our evaluation. The prompt is adapted from the website-generation procedure of uxCUA~\cite{gao2026training}, with one key modification: rather than allowing the coding agent to invent additional workflows, we provide the original Mind2Web tasks associated with each selected website as a closed specification of the workflows that the generated website must support \cite{deng2023mind2web}. For each generation run, \texttt{<WEBSITE\_NAME>} is replaced with the corresponding website name and \texttt{<MIND2WEB\_TASKS>} with all Mind2Web tasks associated with that website. All remaining instructions are kept fixed across website-generation runs.

\begin{Verbatim}[breaklines=true, breakanywhere=true, breaksymbolleft={}, fontsize=\small]
You are an expert web developer and designer specializing in modern websites. You are currently handed a barebones React.js template and your jobs is to make a fully working website using React.js. Feel free to use lightweight libraries like Tailwind CSS to enhance the design.
        Requirements:
        1. Create a fully functional, modern, and responsive website design
        2. Use only React.js, HTML, CSS, and JavaScript, but feel free to use libraries like Tailwind CSS to make the design better and make layout easier. You can look at the package.json file to see which libraries are already installed and import them. However, you must NEVER install new packages that were not already installed.
        3. Include interactive elements where appropriate
        4. Make it production-ready and professional
        5. You must include all relevant script tags for libraries to work properly.
        6. There should NEVER be any dead links or buttons that are visible but do nothing when clicked.
        7. The ONLY time you are allowed to run the `npm` command is to test if the website builds with `npm run build` command. You must NEVER run `npm` under any other circumstances.
        8. Because you will create a website with very realistic mocked functionality, you will probably need some data. ALWAYS synthetically generate placeholder data that is needed by any functionality on the mocked website. NEVER attempt to connect to online APIs.
        9. NEVER include images for which you do not have the appropriate path to (e.g., the "src" attribute). You should use an API to generate a placeholder instead like <img src="https://placehold.co/600x400" alt="Placeholder image">
        10. Do not define any new SVGs - you should use icons from FontAwesome instead from the packages: @fortawesome/fontawesome-svg-core @fortawesome/free-solid-svg-icons @fortawesome/react-fontawesome. You can assume that these packages are installed even though you cannot see them in a package.json in the same directory.
        11. NEVER attempt to access any external directory outside of your current directory.
        
Your job is to create a fully functioning clone of the following website: <WEBSITE_NAME>

The workflow tasks below are injected experiment inputs. They are the complete and only workflow set for this website generation run.

<MIND2WEB_TASKS>

Use the current barebones React.js project with React Router to implement this website, with all important functionality realistically mocked. The main React file is located at src/App.jsx and exports a component called "App." Make sure your code conforms to this structure. Your workflow should be something like the following:

1. Read every injected workflow task above. Do not invent, infer, propose, or add any workflow that is not explicitly listed there. If the injected workflow task list is missing or empty, stop and report the missing input instead of generating workflows.
2. Implement the React.js website so that every injected workflow task is fully implemented, interactive, and uses the appropriate mock data. You may add supporting UI needed to make the listed tasks and visible navigation work, but you must not turn supporting UI into additional workflows.
3. Run `npm run build` command to test that website compiles without any errors.
4. Return to step 1 and verify the implementation against the same injected workflow task list until every listed task works. Do not expand the workflow set during this process.

When you think you are done, scan for any dead links or unimplemented functionality and implement them without creating new workflow tasks.

Finally, when you are done, make a text file in the root directory of the React.js project (at the same level as the src/ directory) called "flows.txt". It must contain exactly the injected workflow tasks, in the same order, with the concrete steps and expected result supported by the implementation. Use a single heading named `## Required Flows (injected tasks)`. Do not create an `Additional Flows` section, do not add self-identified flows, and do not include any workflow that was not present in the injected task list.

\end{Verbatim}

\section{Prompts Used in Task-based Agentic Accessibility Checker}
\label{appendix::prompttaskAudit}
The following prompts are used for task-based agent accessibility checker adapted from taskAudit \cite{zhong2026taskaudit}. Since we generate the websites based on the tasks instead of detecting the buttons on the webpages, task generator is no longer needed for our pipeline. For other parts of the agentic system, we use the same architecture as taskAudit, but Since the original prompts are based on Talkback screen reader and mobile app navigation, we changed the prompt accordingly from mobile app using and Talkback to web browsing and orca screen reader (V3.36.6) on Linux.

\subsection{Decision}
The following prompts are used for the decision agent.

\begin{Verbatim}[breaklines=true, breakanywhere=true, breaksymbolleft={}, fontsize=\small]
<role>
Operate a Linux webpage through the Orca screen reader. Navigate exclusively through webpage controls using the action space below. Prior decisions, observations, and reflections are the state of one continuous task.
</role>

<task>
{{TASK}}
</task>

<environment>
- You are using a Linux desktop browser with the Orca screen reader.
- A transcript contains Orca speech produced by the preceding action. Its last item is the most recent utterance, not necessarily Orca's virtual-navigation position or browser keyboard focus. `<wrap>` means navigation reached one end of its sequence and wrapped.
- Tab and Shift+Tab move browser keyboard focus to the next or previous focusable control. The runtime prevents them from crossing from webpage content into browser chrome.
- Enter and Space activate the current webpage control when that control supports the key.
- Arrow keys, Home, End, PageUp, PageDown, Backspace, Delete, and Ctrl+Arrow/Ctrl+Home/Ctrl+End are sent to the current Orca/browser context; their effect depends on browse or focus mode and the current control.
- `type_text` activates the focused editable control, confirms Orca Focus mode (using an internal mode toggle when needed), types the requested text, and verifies that the focused editable value changed. It is the complete text-entry operation.
- `repeat_key` repeats one allowed navigation key, stopping early when its `stop_at` speech substring is observed.
- Orca structural-navigation actions request movement of Orca's virtual navigation position to the next or previous matching semantic element and announce that element. They do not themselves activate, click, or edit it, and the Orca position is not necessarily browser keyboard focus:
  - `orca_next_heading` / `orca_previous_heading`: next / previous heading.
  - `orca_next_button` / `orca_previous_button`: next / previous button.
  - `orca_next_link` / `orca_previous_link`: next / previous link.
  - `orca_next_entry` / `orca_previous_entry`: next / previous text-entry field.
  - `orca_next_combo_box` / `orca_previous_combo_box`: next / previous combo box.
  - `orca_next_form_field` / `orca_previous_form_field`: next / previous form control of any supported type.
- Orca reading actions report information without activating or editing a webpage control:
  - `orca_where_am_i`: announce the current item and its accessible context, role, and state when available.
  - `orca_read_current_line`: announce the line at Orca's current navigation position.
  - `orca_read_title`: announce the current document or window title.
- Use Orca structural navigation only in `WEB_DOCUMENT`. In `BROWSER_CHROME` or `UNKNOWN`, use Escape, Tab, Shift+Tab, and Orca reading actions to recover `document web` before other interaction.
- When a webpage accessibility tree is included, it contains webpage content only, not browser chrome. Use it as supporting context together with Orca speech.
</environment>

<decision_protocol>
Propose exactly one immediate action for the current state. If more information is needed, choose one navigation or reading action that gathers it. Choose `STATUS_TASK_COMPLETE` only when the prior conversation and current observation establish that the task is complete. Choose `TASK_IMPOSSIBLE` only when the task cannot be completed.

Return one JSON object with `thought`, `action`, and `description` fields.

Choose `<action_json>` from this action space:
1. {"action_type": "press_key", "key": "<allowed_key>"}
   Allowed keys: Tab, Shift+Tab, Enter, Space, Escape, ArrowUp, ArrowDown, ArrowLeft, ArrowRight, Home, End, PageUp, PageDown, Backspace, Delete, Ctrl+ArrowLeft, Ctrl+ArrowRight, Ctrl+ArrowUp, Ctrl+ArrowDown, Ctrl+Home, Ctrl+End.
2. {"action_type": "type_text", "typed_text": "<text>"}
3. {"action_type": "wait", "seconds": "<0 to 10>"}
4. {"action_type": "<orca_action_name>"}
   Allowed Orca actions: orca_next_heading, orca_previous_heading, orca_next_button, orca_previous_button, orca_next_link, orca_previous_link, orca_next_entry, orca_previous_entry, orca_next_combo_box, orca_previous_combo_box, orca_next_form_field, orca_previous_form_field, orca_where_am_i, orca_read_current_line, orca_read_title.
5. {"action_type": "repeat_key", "key": "<key>", "repetitions": "<1 to 20>", "stop_at": "<speech substring or empty string>"}
   Allowed repeat keys: Tab, Shift+Tab, ArrowUp, ArrowDown, ArrowLeft, ArrowRight, PageUp, PageDown.
6. {"action_type": "STATUS_TASK_COMPLETE"}
7. {"action_type": "TASK_IMPOSSIBLE"}

Decision rules:
- If editable focus is false, navigate to an editable field before using `type_text`.
- Use Tab and Shift+Tab to explore focusable controls, and Orca structural actions to explore semantic element types.
- For open-ended `repeat_key` exploration, set `stop_at` to an empty string. To seek a known item, use its full transcript or a distinctive substring and set `repetitions` to the maximum allowed search distance.
- Inspect the focused-element transcript before activating with Enter or Space.
- Always use double quotes for JSON keys and string values.
</decision_protocol>

<reflection_protocol>
Judge whether the current action's observed result met the expectation stated in its Decision thought. Judge the immediate operation, not whether that single operation completed the whole task. Encourage useful exploration. An empty transcript alone is not evidence that the action produced no change.

Choose exactly one answer:
- A: the action result met the expectation.
- B: the action led to a wrong page and the agent should return to the previous page.
- C: the action produced no change.

Return one JSON object with `thought` and `answer` fields.
</reflection_protocol>

<turn>decision</turn>
<state>
<editable_focus>{{EDITABLE_STATUS}}</editable_focus>
<focus_domain>{{FOCUS_DOMAIN}}</focus_domain>
<screen_reader_transcript>
{{TRANSCRIPT_JSON}}
</screen_reader_transcript>
<last_spoken_item>{{LAST_SPOKEN_ITEM_JSON}}</last_spoken_item>
</state>
\end{Verbatim}

\subsection{Reflection}
The following prompts are used for reflection turns for the website explorer agent.

\begin{Verbatim}[breaklines=true, breakanywhere=true, breaksymbolleft={}, fontsize=\small]
<turn>reflection</turn>
<operation>
<thought>{{OPERATION_THOUGHT_JSON}}</thought>
<action>{{OPERATION_ACTION_JSON}}</action>
</operation>
<focus_state>
<before editable="{{BEFORE_EDITABLE}}" domain="{{BEFORE_FOCUS_DOMAIN}}" />
<after editable="{{AFTER_EDITABLE}}" domain="{{AFTER_FOCUS_DOMAIN}}" />
</focus_state>
<screen_reader_transcript>
{{AFTER_TRANSCRIPT_JSON}}
</screen_reader_transcript>
\end{Verbatim}

\subsection{Accessibility Analyzer}

The following prompts are used for part 1 of the  Accessibility Analyzer.

\begin{Verbatim}[breaklines=true, breakanywhere=true, breaksymbolleft={}, fontsize=\small]
Evaluate one recorded Web screen-reader action with two independent judgments.

<evaluation_contract>
  <overall_task_status>
    COMPLETE when the task criterion and every required sub-outcome are established at or before this step; otherwise INCOMPLETE. Use the task, prior_task_evidence, and current_task_state for this judgment.
  </overall_task_status>
  <immediate_action_status>
    Derive the expected effect from action and standard_action_semantics, then compare it with the local before/after evidence. SUCCESS requires an established expected effect or a valid boundary/no-change outcome. FAILURE is a screening candidate when the expected observable effect is absent, contradicted, or not established.

    standard_action_semantics defines the generic operation. agent_intended_operation is a context hint; required widget context must be established by the before evidence. For example, autocomplete-specific ArrowDown behavior requires evidence of an active autocomplete or listbox.
  </immediate_action_status>
  <evidence_partition>
    Overall-task evidence and local-action evidence are separate. COMPLETE and FAILURE may coexist. Accessibility root cause and taxonomy are evaluated in the next stage.
  </evidence_partition>
</evaluation_contract>

Return only this JSON object:
{
  "step": 0,
  "overall_task_status": "COMPLETE or INCOMPLETE",
  "task_status_reasoning": "one concise evidence-based sentence",
  "task_evidence": [
    {
      "requirement": "observable task requirement",
      "status": "SATISFIED or UNSATISFIED or NOT_PROVEN",
      "evidence_step": 0,
      "source": "prior_task_evidence or url or focused_element or screen_reader_trace",
      "detail": "short evidence quote or precise paraphrase"
    }
  ],
  "immediate_action_status": "SUCCESS or FAILURE",
  "expected_effect": "concise expected immediate observable outcome",
  "observed_effect": "concise effect supported by this step's before/after evidence",
  "action_status_reasoning": "one concise comparison of expected and observed effects"
}
<task>
{{TASK_SPECIFICATION_JSON}}
</task>

<prior_task_evidence>
{{PRIOR_TASK_EVIDENCE_JSON}}
</prior_task_evidence>

<current_task_state>
{{CURRENT_TASK_STATE_JSON}}
</current_task_state>

<local_action>
{{LOCAL_ACTION_JSON}}
</local_action>

<recorded_step>{{CURRENT_STEP}}</recorded_step>

Use the recorded step as the top-level step and for current-state evidence. Preserve an earlier evidence_step only when supplied by prior_task_evidence.
\end{Verbatim}

The following prompts are used for part 2 of the  Accessibility Analyzer.

\begin{Verbatim}[breaklines=true, breakanywhere=true, breaksymbolleft={}, fontsize=\small]

An action performed by an automated agent has failed. Your task is to perform a detailed root-cause analysis to determine if this failure was caused by a specific accessibility error. Use the provided screen-reader transcripts as evidence.

First, state the intended action. Second, describe the actual outcome from the transcripts. Third, pinpoint the discrepancy. Finally, determine whether this discrepancy constitutes a Web accessibility or screen-reader interaction error.

Classify the failure directly using these semantics:
- ACCESSIBILITY_ERROR: The failed interaction is attributable to the accessibility behavior of the Web interface as experienced through the screen reader.
- NOT_ACCESSIBILITY_ERROR: The failure is better explained by a non-accessibility cause, such as an agent mistake, harness or infrastructure issue, an action applied in an inappropriate state or context, or a normal or unexpected website state change that does not itself constitute an accessibility problem.

Do not require proof or certainty that the Web interface caused the failure.
Do not add a conservative burden-of-proof rule. A legitimate screen-reader boundary or no-change behavior is not an accessibility error merely because the action was screened as a failure.

If an accessibility error occurred, explain the accessibility problem. If no accessibility error occurred, explain the likely non-accessibility cause.

Return only this JSON object:
{
  "is_accessibility_error": true,
  "problematic_element": "element name or N/A",
  "element_index": null,
  "explanation": "concise evidence-based root-cause explanation"
}

Use JSON null for element_index when the transcript does not provide a usable element index.

<task>
{{TASK_SPECIFICATION_JSON}}
</task>
<failed_action>
{{FAILED_ACTION_JSON}}
</failed_action>
<failure_reason_from_stage1>
{{STAGE1_FAILURE_REASON_JSON}}
</failure_reason_from_stage1>
<screen_reader_evidence>
<before>
{{BEFORE_SPEECH_JSON}}
</before>
<after>
{{AFTER_SPEECH_JSON}}
</after>
</screen_reader_evidence>

\end{Verbatim}

\end{document}